\documentclass[prc,tightenlines,nofootinbib,superscriptaddress]{revtex4-2}

\usepackage{tipa}
\usepackage{mathrsfs}
\usepackage{amsmath,calc,amsfonts,amssymb}
\usepackage{graphicx}
\usepackage[dvipsnames]{xcolor}
\usepackage{tikz}
\usepackage{subcaption}
\usepackage[T1]{fontenc}
\usepackage[english]{babel}
\usepackage[utf8]{inputenc}
\usepackage{soul}
\usepackage{xfrac}
\usepackage{bm}	
\usepackage[colorlinks=true,linkcolor=Blue,urlcolor=Blue,citecolor=Red]{hyperref}
\usepackage{cleveref}
\usepackage[top=0.5in,bottom=0.5in,left=0.7in,right=0.7in]{geometry}
\usepackage{bbold}
\usepackage{mathtools}
\usepackage[mathscr]{euscript}
\usepackage[title]{appendix}

\numberwithin{equation}{section}

\renewcommand{\theequation}{\thesection.\arabic{equation}}
\renewcommand{\theequation}{\thesection.\arabic{equation}}
\newcommand\this{\addtocounter{equation}{1}\tag{\theequation}}
\newcommand{\eq}[1]{\begin{align*}#1 \end{align*} }
\newcommand{\eqn}[1]{\begin{align}#1 \end{align} }
\newcommand{\e}{\text{e}}
\newcommand{\half}{\frac{1}{2}}

\DeclareMathOperator{\sech}{sech}

\newcommand{\be}{\begin{eqnarray}}
\newcommand{\ee}{\end{eqnarray}}

\newcommand{\gsim}{\mathrel{\hbox{\rlap{\lower.55ex \hbox {$\sim$}}
			\kern-.3em \raise.4ex \hbox{$>$}}}}
\newcommand{\lsim}{\mathrel{\hbox{\rlap{\lower.55ex \hbox {$\sim$}}
			\kern-.3em \raise.4ex \hbox{$<$}}}}
\newcommand{\T}[1]{\Theta\left(#1\right)}
\newcommand{\zot}{\frac{z}{t}}
\newcommand{\mat}{\text{Mat.}}
\newcommand{\rad}{\text{Rad.}}

\begin{document}
	
	\title{ Distribution of Nuclear Matter and Radiation in the Fragmentation Region }
	
	\author{Isobel Kolb\'e}
	\email{ikolbe@uw.edu}
	
	\affiliation{Institute for Nuclear
		Theory, University of Washington, Box 351550, Seattle, WA, 98195, USA  }
	\author{Mawande Lushozi}
	\email{mlushozi@uw.edu}
	\affiliation{Institute for Nuclear
		Theory, University of Washington, Box 351550, Seattle, WA, 98195, USA  }
	
	\author{Larry D. McLerran}
	\email{mclerran@me.com}
	\affiliation{Institute for Nuclear
		Theory, University of Washington, Box 351550, Seattle, WA, 98195, USA  }

	\author{Gongming Yu}
	\email{ygmanan@uw.edu}
	\affiliation{Institute for Nuclear Theory, University of Washington, Box 351550, Seattle, WA, 98195, USA }
	\affiliation{CAS Key Laboratory of High Precision Nuclear Spectroscopy and Center for Nuclear Matter Science, Institute of Modern Physics, Chinese Academy of Sciences, Lanzhou 730000, China}
	\affiliation{Harbin Engineering University, Harbin 150 000, China}
	
	
	
	
	

	\begin{abstract}
		We study the fragmentation (far forward/backward) region of heavy ion collisions by considering an at-rest nucleus which is struck by a relativistic sheet of colored glass. By means of a simple classical model, we calculate the subsequent evolution of baryons and the associated radiation.  We confirm that the struck nucleus undergoes a compression and that the dynamics of the early times of the collision are best described by two separate fluids as the produced radiation's velocity distribution is very different to the velocity distribution of the matter in the struck nucleus.
		
	\end{abstract}

	\maketitle



    \section{Introduction}


	The fragmentation (very forward/backward) region of heavy-ion collisions is interesting as it is the only baryon-rich \cite{Li:2018ini} part of phase-space in the presence of strong gluon fields. 
	It was first observed in \cite{Anishetty:1980zp}  that, upon interaction with a relativistic heavy-ion projectile, the target will undergo some compression and that this in turn will increase the baryon and energy density in the fireball.  A fully quantum calculation of the distribution of baryons in the fireball was done in \cite{McLerran:2018avb}, but the result is difficult to interpret at early times.   By means of a classical model of the target we would like to achieve two goals:  the first is to reevaluate the compression of the nucleus to include the effects of saturation in the framework of the color-glass condensate (CGC) (see for eg.\, \cite{Kovchegov:2012mbw,Iancu:2003xm,Weigert:2005us});  and our second goal is to develop an intuitive picture of the early-time dynamics of both the struck particles and the resultant radiation. 
	
	Our model of the target assumes the nucleus is made of uniformly distributed constituent quarks with negligible interactions. We will focus on the $z$-component (beam direction) of the evolution of the distribution of quarks and radiation so that the majority of the calculation is performed in (1+1)D. We will work in the target's rest frame\footnote{In contrast to the approach followed in \cite{Kovner:1995ja} which deals with the problem in the center of mass frame and computes the radiation that results from colliding two infinitesimally thin sheets.  That approach naturally yields a longitudinally invariant gluon distribution, which is not the case in the present work since we do not have a longitudinally invariant initial distribution of matter.  As such, the two results are not related by a simple Lorentz transformation.}
	where the collision will be a sequence of independent successive interactions of a sheet of colored glass with each quark. We will treat these interactions entirely classically, and so each quark follows the trajectory of a free classical particle after it is struck.
	
	The problem of a single classical color-charged point particle interacting with a sheet of colored glass has recently been considered \cite{Kajantie:2019hft,Kajantie:2019nse}, where it was discovered that the quark recoil needs to be taken into account when the transverse momentum of the radiated gluon is high.
	  In the current work we will use the classical quark-CGC interaction of Ref. \cite{Kajantie:2019hft,Kajantie:2019nse} to treat the evolution of the quarks while approximating the gluon radiation spectrum by a flat distribution. This approximation aligns well with experiment, which shows a slowly varying rapidity distribution \cite{Adam:2016ddh}, and suits our purposes better than the more detailed first-principles treatments found in the literature \cite{Kajantie:2019hft,Kajantie:2019nse,Kovchegov:1998bi,Kopeliovich:1998nw,Kopeliovich:1999am,Lushozi:2019duv}.    
	
	Some attempts have been made to study the hydrodynamical evolution of the forward-moving baryon-rich fluid in a heavy-ion collision \cite{Gyulassy:1986fk,Li:2018fow}.  
	Our results will have implications for the use of (3+1)D, baryon-rich hydrodynamical transport equations \cite{Shen:2018pty,Schenke:2010nt,Denicol:2018wdp,Du:2019obx} in the fragmentation region as we will argue that the two fluids (the baryon-rich nucleonic matter and the baryon-free produced matter) have very different velocities and do not equilibrate in the early stages of the collision.  An interesting prospect would be to apply models such as three-fluid hydrodynamics \cite{Ivanov:2005yw,Ivanov:2018rrb} to the fragmentation region.

	This paper is organized as follows: In  \cref{app:GluonRad} we present a brief review of the equations of motion describing a classical particle interacting with a sheet of colored glass.  We then use these equations of motion to derive a momentum distribution of the struck quark after averaging over the classical sources.  We will use the McLerran-Venugopalan (MV) model \cite{McLerran:1993ka} so that the averaging is performed over a Gaussian distribution of sources.  
	
	In \cref{sec:f_dN_dE} we start to develop our model by deriving the phase-space densities for the struck nucleons (hereinafter the ``matter'' particles) as well as the phase space density for the resulting radiation (hereinafter the ``radiation'').  Armed with the phase-space density, we then compute the number- and energy densities in a manner that is straightforward but yields interesting physics.
%
	
	
%

	We then move on to compute the average velocity and, by extension, the average momentum-space rapidity of both the matter and the radiation in \cref{sec:AvgVandY}.  It is the results in this section that allow us to come to an important realization;  We find that the matter and the radiation behave very differently at early times, suggesting that the a full, hydrodynamical treatment of the problem requires initial conditions that contain the physics of two fluids with very different properties that do not equilibrate soon after the collision.
	
	We summarize and discuss the above results in \cref{sec:SummaryAndDiscussion} where we highlight the realization of our two goals.

    \section{ Review: Momentum-kick of a classical particle  interacting with a sheet of colored glass}\label{app:GluonRad}

    Let us review some results from \cite{Kajantie:2019hft,Kajantie:2019nse} describing the interaction of a sheet of colored glass with a classical point particle. In the MV model \cite{McLerran:1993ka}, gluon fields, generated by classical color sources, are governed by the classical Yang-Mills equation 
    \eqn{D_\mu\mathbf{F}^{\mu\nu}&=\mathbf{J}^{\nu} \label{YangMillsEqnVec}
    }
    Where the current associated with a sheet of colored glass is:
    \eqn{J_a^{\mu}&=\delta^{\mu +} \delta(x^-) \rho_a(x_\perp)\,\, .
    }
    The 4-momentum of a point charge interacting with this sheet evolves according to 
    \eqn{\frac{dp^\mu}{d\tau}=g \mathbf{T}\cdot \mathbf{F}^{\mu\nu}u_\nu \,, \label{p-eqn-o-m}
    } 
    where $\mathbf{T}$ is the classical color charge vector associated with the point particle. In the $A^+$ gauge, we have  $F^{+i}= -\partial^{i}A^{+}$ and so the ``-'' component of \cref{p-eqn-o-m} has the simple form 
    \eqn{ \frac{d p^-}{d\tau}=0.\label{pminus-const}	
    }
     The above equation \cref{pminus-const} tells us that $p^-$ is unchanged by the interaction so 
    \eqn{ p_{\text{\tiny{before}}}&=(m/\sqrt{2},m/\sqrt{2},0_{\perp})\label{p-before}\\
    	p_{\text{\tiny{after}}}&=(p^+,m/\sqrt{2},\mathbf{p}_\text{\tiny{T}})\,\,.\label{p-after}	
    }
    This then means $\mathbf{p}_\text{\tiny{T}}$  and $p^z$ after the collision are related.  To see this, note that imposing the mass-shell condition on \cref{p-after} gives 
    \eqn{p^+=\frac{m^2+p_{\text{\tiny{T}}}^2}{\sqrt{2}m}
    }
    	\eqn{\therefore p^z =\frac{p^+-p^-}{\sqrt{2}}	=\frac{p_{\text{\tiny{T}}}^2}{2m}\,\,. \label{pT-pz}
    }

    It also follows from \cref{p-eqn-o-m,pminus-const} that the particle gets a transverse momentum-kick 
     \eqn{\frac{\partial p^i}{\partial x^-} &=gT\cdot F^{+i}\,\,,   
     }
    so that
    	\begin{align}
    		p^i(x_\perp)&=gT^a\frac{\partial^{i}}{\nabla_T^2} \rho^a(\vec{x}_\perp)\,\,.\label{ML1}
    	\end{align}

    According to the MV model the charge density $\rho^a$ is sampled from a gaussian distribution
    	\begin{align}
    		W_{[\rho]} &=\frac{1}{\mathscr{N}}\exp\left\lbrace-\frac{1}{2\mu_A^2}\int d^2\!x\rho^a(\vec{x}_{ _\perp}) \rho^a(\vec{x}_{ _\perp})\right\rbrace.
    	\end{align}
    
    We can therefore extract the following probability distribution over~$\vec{p}_{ _T}$
    	\begin{align}
    		\frac{d P(\vec{p}_{ _T})}{d^2 p_{ _T}}&= \frac{2}{(gT^a)(gT^a)\mu_A^2\ln\frac{Q_s}{\Lambda} }\exp\left(-\frac{2\pi p_T^2}{(gT^a)(gT^a)\mu_A^2\ln\frac{Q_s}{\Lambda} }\right)\,\,.
    	\end{align}
    Since we have established that $p^z$ and $p_{\text{\tiny{T}}}$ are related, we have that  
    \eqn{\frac{dP(p^z)}{dp^z}&=\frac{1}{p_0}e^{-\frac{p^z}{p_0}}\,\,
    }
    where
    \eqn{p_0=\frac{(gT^a)(gT^a)\mu_A^2\ln\left(\frac{Q_s}{\Lambda_{\text{\tiny{QCD}}}}\right)}{4\pi m}\simeq Q_s^2/2m\,\,.
    } 
    	
      We have used a definition of $Q_s$ consistent with \cite{Iancu:2003xm}\footnote{Following \cite{Iancu:2003xm} we interpret $(gT^a)(gT^a)$ the square of the classical color charge carried by the particle, given by the quadratic Casimir $(gT^a)(gT^a)=g^2C_R$ of the chosen representation. The saturation scale as defined in \cite{Iancu:2003xm} involves gluons and so the adjoint representation ($C_R=C_A=N_c$) is used. For our purposes it only matters that the charges are classical and so we leave the exact representation ambiguous.}. 
      One may estimate the value of $Q_s$ \cite{McLerran:2018axu} in the fragmentation region at RHIC and the LHC to find that $Q_s\sim 1-2.5$ GeV at RHIC and $Q_s\sim 2.5-7$ GeV at the LHC.  The details of this derivation have been relegated to \cref{app:pT-dist}.

\section{Phase-space, number density, and energy density}\label{sec:f_dN_dE}

	The fundamental object we will deal with is the phase-space density which we will first derive separately for the nucleons (or quarks) and the radiation.  We will then compute the number- and energy densities directly from the phase-space density.
	
\begin{figure}
			\centering
			\includegraphics[scale=1]{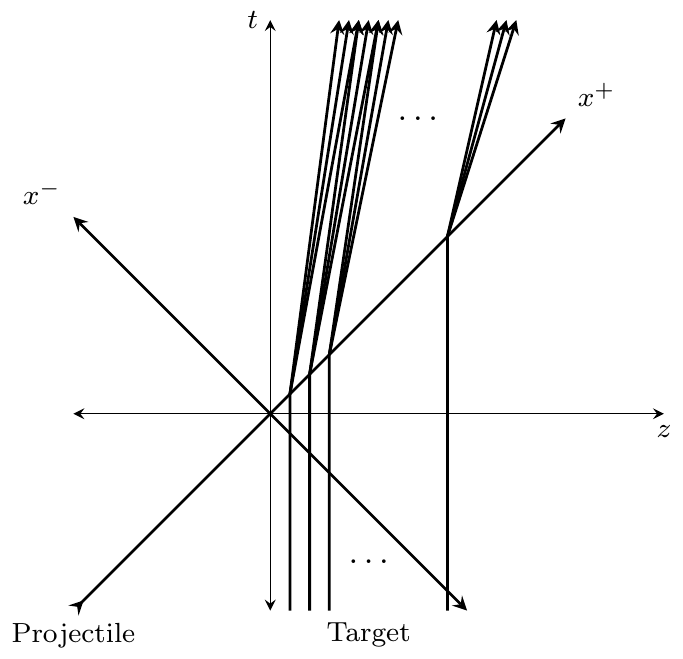}
			\caption{Longitudinal kinematics of the process in (1+1)D, a string of stationary quarks that are struck by a relativistic ($ v=1 $) sheet of colored glass, causing the quarks to move off with varying momenta. \label{fig:Kinematics}}
		\end{figure}

	We are interested in the single particle distribution for quarks inside a nucleus that is struck by a sheet of colored glass. In (1+1)D, we consider $N$ uniformly distributed quarks of the target nucleus, arranged in a line of length $R$.  In the target's rest frame, the quarks all have zero initial velocity and are then struck, one after the other, by a sheet of colored glass  moving along $x^{-}=0$. The struck quarks will then radiate additional particles. The geometry of the process is shown in \cref{fig:Kinematics}.
		
	\subsection{Phase-space: Radiation}

		Since we will employ a simple model for the production of the radiation, it is pedagogically simpler to consider the radiation first.  We will consider a situation in which the production distribution of the radiation is flat as a function of the rapidity $ F(y) =F$, and the multiplicity is ``fitted'' to experimental data.  The flatness assumption is justified by the experimental observation that the meson distribution varies slowly with pseudo-rapidity \cite{Adam:2016ddh}. We assume a picture where the radiated particles are produced at the initial positions of the quarks, $ z \in (0,R) $, at the time each of the quarks is hit by the sheet.  
		
		
		One may include the effects of a formation time, but the additional complexity reduces the clarity of the analytic results.  One may also consider a limit to the beam rapidity, $ y_b $.  In practice this limit serves as a useful regulator, but it has little effect on the final result.  At LHC energies, the beam rapidity (in the rest-frame of the target nucleus) is around 16, so that $ \tanh y_b\sim 1 $.    Since the result is analytical and easily interpreted, we will present here only the derivation without formation time or beam rapidity limit.  The density distribution of the radiation with formation time is derived in full in  \cref{App:RadDeriv}.
		
		We start by considering the phase-space density of the radiation produced by a single struck particle, situated at the origin:		
			\begin{equation}
				f^{0}(z,t,y) = F\, \T{t}\,\delta(z-t\tanh y).
			\end{equation}
			
		A single particle at an arbitrary position $ z_i $ will radiate radiation with with a phase-space density given by
			\begin{equation}
				f^{z_i} = F\, \T{t-z_i}\,\delta(z-z_i-(t-t_i)\tanh y).\label{eq:firad}
			\end{equation}
		
		Suppose now that the total number of quarks $ N $ is distributed along $ (0,R) $ uniformly so that there are  $N_i= \sfrac{N}{R} $ quarks at each $ z_i $, each of which radiates radiation with phase-space density given by \cref{eq:firad}.  The full phase-space density is then given by
			\begin{equation}
				f(z,t,y)=\sum_i f^{z_i}(z,t,y).
			\end{equation}
		
		We may go to the continuum limit, taking careful note of the fact that the integral over quark positions needs to be split in order to ensure that only those locations that contain at least one quark are integrated over.  The phase space density of the radiation from an at-rest nucleus struck by a relativistic sheet of colored glass is then given by:
			\begin{align}
				f(z,t,y)	
					=&\frac{F\,N}{R}\Bigg(\T{R-z}I^{\rad}_{f}(z)
						+\T{z-R}I^{\rad}_{f}(R)\Bigg),\label{eq:fzty_rad}
			\end{align}
		
		where
			\begin{align}
				I^{\rad}_{f}(a)
					=&\int_0^a dz'\T{t-z'}\delta(z'(\tanh y-1)+z-t\tanh y)\nonumber\\
					=&\frac{1}{1-\tanh y}\T{z-t \tanh y}\T{t \tanh y-z-a(\tanh y-1)}.
			\end{align}
		
		We will need to make some numerical estimates: A central lead-lead collision at the LHC typically produces on the order of 2 000 charged particles per unit of pseudo-rapidity \cite{Adam:2015ptt,Abbas:2013bpa}, spanning about 10 units of pseudo-rapidity, and, since the total number of particles produced (including neutral particles) is around twice that, one expects that around 40 000 particles are produced in such collisions.  We can then estimate that the amount of radiation per valence quark to be around $ 40\, 000/400/3\sim30 $, so that we will take  $ F\sim 30. $

%
	
	\subsection{Phase-space: Matter}\label{Matt-dist}
	
		In deriving the phase-space density of the matter distribution, we will follow a similar process as for the derivation of the radiation, but we will make sure to incorporate the physics of \cref{app:GluonRad} in the momentum distribution. Assuming no interaction among the quarks, one arrives at the following single particle distribution for the matter, the derivation of which is described in \cref{app:Deriv_matter}:
			\begin{align}
			f(x,p)
			=&
			\frac{N}{R}\theta(t-z)\theta\left(z-\frac{tp}{p+m}\right)\nonumber\\
			&\qquad\times\theta\left(\frac{p(t-R)}{p+m}+R-z\right)\left(1+\frac{p}{m}\right)\frac{1}{p_0}e^{-p/p_0}\nonumber\\
			&+\frac{N}{R}\theta(z-t)\theta(z)\theta(R-z)\delta\left(p\right),\label{f-x-p_matter}
			\end{align}
		
		where $ \theta $ is the usual Heaviside-theta function, $x=(t,z)$, $m$ is the quark mass, $p_0$ is the typical momentum given to a struck quark, and we have suppressed the $ z $-superscript so that $p$ is the $z$-component of the quark's momentum\footnote{From here on we will continue to suppress the superscript on $p^z$.}. 
		
		The distribution $f(x,p)$ of \cref{f-x-p_matter} above has two terms; one term with the prefactor $\theta(t-z)$ and another term with $\theta(z-t)$, describing the distribution of the quarks behind the sheet and those ahead of the sheet, respectively. One can see that before the collision, $t<0$, the particles are described entirely by the $\theta(z-t)$-term, involving only a $\delta(p)$ which describes a distribution of stationary particles. As the sheet progresses through the nucleus, the quarks behind the sheet are left with a momentum distribution that depends on the physics of \cref{app:GluonRad}, until the sheet has passed through the nucleus ($ t>R $), whereafter the distribution is described entirely by the momentum distribution of struck quarks. The theta function $\theta(z-pt/(p+m))$ behind the sheet, simply means that a particle at $(z,t)$ has a velocity($z$-component) less than $z/t$. This makes sense since, $z/t$ is the velocity of a quark that was struck at $z=0,t=0$, and particles located at $(z,t)$ are all those that the initial quark from the origin has caught up with, all of which must obviously be slower moving. 
		The remaining theta function of the behind-the-sheet term similarly sets a minimum velocity for motion beyond the initial bounds of the nucleus, i.e $z>R$.

		We will make a number of numerical estimates: we will consider a lead nucleus as our target nucleus which then has a diameter (not radius) of $ R\sim 2\times1.1\,\text{fm}\times A^{\sfrac{1}{3}}\sim 14 $\,fm; we will consider $ N=600 \sim 3 \times 208$ quarks in our target nucleus, each quark with a mass of $ 300 $\,Mev $ \sim 938 $\,Mev\,/(3 valence quarks in the proton).  The parameter $ p_0 $ scales like $ Q_s^2/m $.  We will take $ Q_s\sim 1 $\,Gev (corresponding to RHIC energies) so that $ \sfrac{p_0}{m} \sim\sfrac{Q_s^2}{m^2}\sim\sfrac{(1 \text{GeV})^2}{(300\text{MeV})^2}\sim10$.

	\subsection{Number density}\label{sec:dN}
	
		Armed with the phase-space density, one may straightforwardly compute the spatial distribution of the matter and the radiation at a fixed time by integrating the momentum dependence:
			\begin{equation}
				\frac{dN}{dz}=\int_{0}^{\infty}dp\, f(x,p).
			\end{equation}		
			
		The spatial distribution of quarks at a fixed time, $t$, is therefore given by integrating the momentum dependence of \cref{f-x-p_matter}, giving
			\begin{align}
				\frac{dN^{\mat}}{dz}
					&=\int_{0}^{\infty}dp f(x,p)\nonumber\\
					&= \frac{N}{R}\theta(t-z)\left[I_0^{\mat}\left(\frac{z}{t}\right)-\theta(z-R)I_0^{\mat}\left(\frac{z-R}{t-R}\right) \right]+\frac{N}{R}\theta(z-t)\theta(R-z)\theta(z)\,\,,\label{eq:MattDistzt}
			\end{align}
		where
		\eqn{I_0^{\mat}(w)&=-\left(\frac{w}{1-w}+1+\frac{p_0}{m} \right)\exp\left(-\frac{m}{p_0}\frac{w}{1-w} \right)+1+\frac{p_0}{m}\,\,.
		}
		
	
		
		Changing variables to $\eta=\frac{1}{2}\log\left(\frac{t+z}{t-z}\right)$ and $\tau^2=t^2-z^2$, see \cref{app:DeriveEtaTau}, one finds

			\begin{align}
				\frac{dN^{\mat}}{d\eta}
					=& \frac{dz}{d\eta}\frac{dN^{\mat}}{dz}\nonumber\\
					=& \frac{N}{R}\frac{\tau}{\cosh\eta} \theta(\tau)\left[1+\frac{p_0}{m}-\frac{1}{2}\left(\frac{2p_0}{m}+e^{2\eta}+1 \right)\exp\left(-\frac{m}{2p_0}(e^{2\eta}-1) \right)\right.\nonumber\\
					&\qquad\left. -\theta(\tau \sinh\eta-R)\left\lbrace 1+\frac{p_0}{m}-\frac{1}{2}\left(\frac{2p_0}{m}+e^{2\eta}+1-2Re^\eta/\tau \right)\exp\left( -\frac{m}{2p_0}(e^{2\eta}-1-2Re^\eta/\tau)\right) \right\rbrace \right]\nonumber\\
					&+\frac{N}{R}\frac{\tau}{\cosh\eta}\theta(-\tau)\theta(R-\tau\sinh\eta)\theta(\tau\sinh\eta). \label{matt-dist}
			\end{align}

		The distribution of the radiation is similarly obtained by integrating \cref{eq:fzty_rad}:
			\begin{align}\label{eq:dNdzRad}
			\frac{dN^{\rad}}{dz}
				&=\int_{0}^{\infty}dy\,f(z,t,y)\nonumber\\
				&=\frac{F\,N}{R}\T{t-z}\Bigg(\T{R-z}I^{\rad}_{N}(0)
					+\T{z-R}I^{\rad}_{N}\left(\frac{z-R}{t-R}\right)\Bigg),
			\end{align}
		
		where
			\begin{align}\label{eq:dNdzRad2}
			I^{\rad}_{N}(a)
				&=\int_{\tanh^{-1} a}^{\tanh^{-1} \zot}\,dy\frac{1}{1-\tanh y}\nonumber\\
				&=\frac{1}{4}\left(2\tanh^{-1}\sfrac{z}{t}-2\tanh^{-1}a-\e^{2\tanh^{-1}a}+\e^{2\tanh^{-1}\sfrac{z}{t}}\right).
			\end{align}

\begin{figure*}
				\centering
				\begin{subfigure}[b]{0.5\linewidth}
					\includegraphics[scale=0.45]{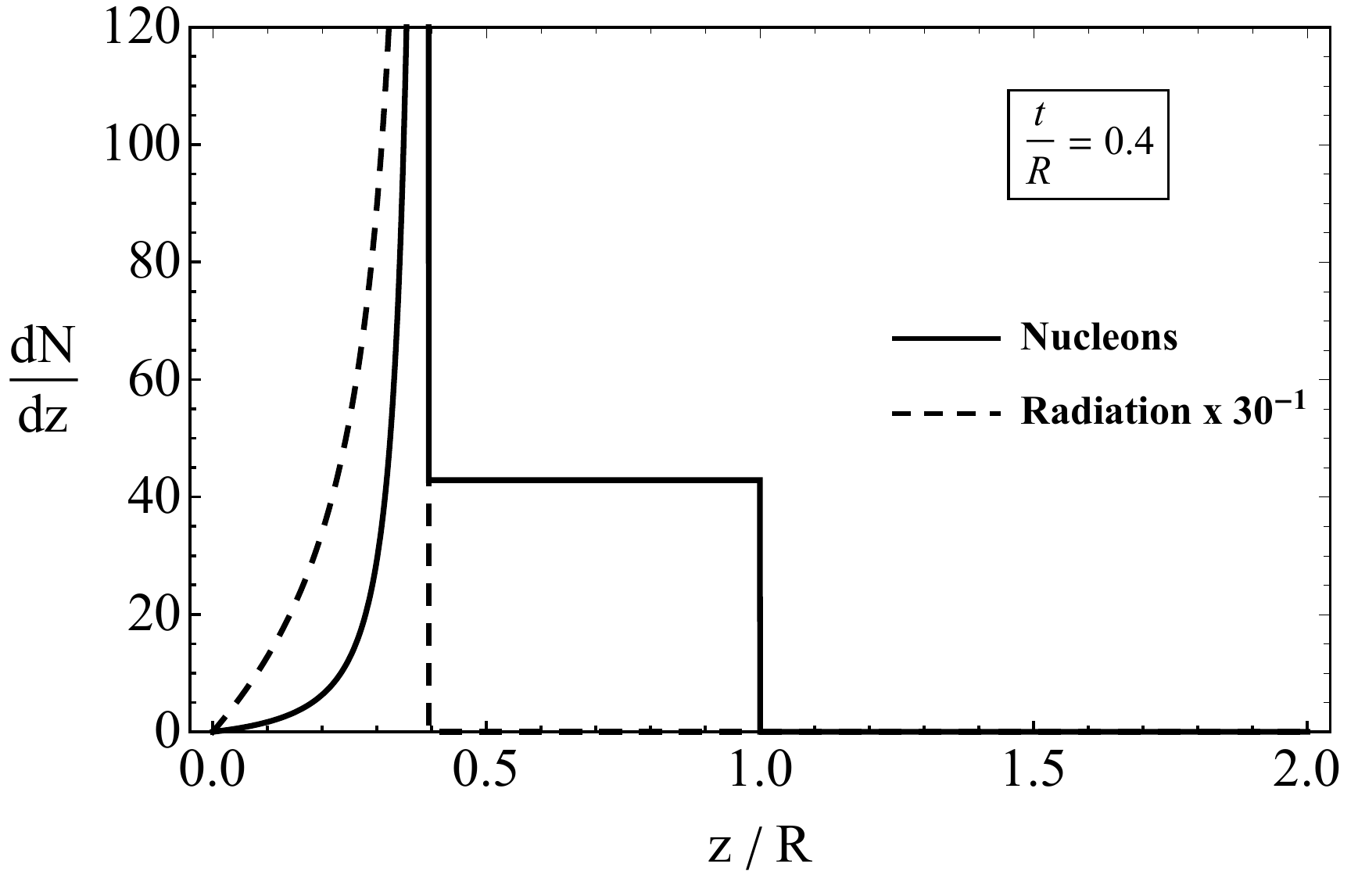}
					\caption{\label{fig:dNdz_EarlyTimes}}
				\end{subfigure}
				\begin{subfigure}[b]{0.5\linewidth}
					\includegraphics[scale=0.45]{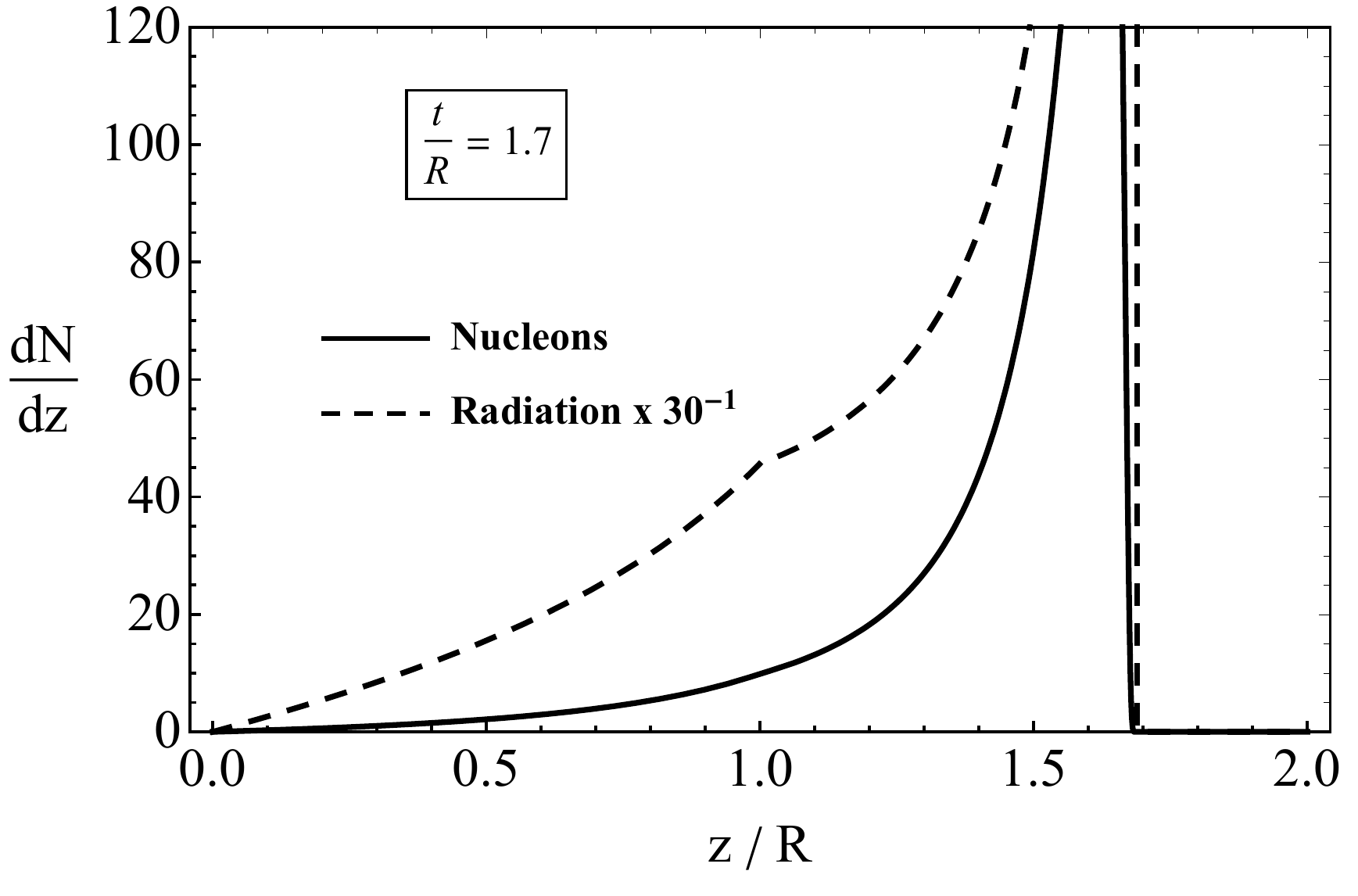}
					\caption{\label{fig:dNdz_MidTimes}}
				\end{subfigure}
				\caption{Distribution of matter and radiation (without formation time) when the shockwave has (a) traversed part of the stationary matter distribution, and (b) moved beyond the initial matter distribution.  Both figures use $ N= $ 600 matter particles, $ \frac{p_0}{m}= 10$, and $ F= $ 30.\label{fig:dNdz_EarlyTimes_MidTimes}}
			\end{figure*}

		We may, once again, do a coordinate transformation to $ (\eta,\tau) $-space (see \cref{app:DeriveEtaTau}), but we will not reproduce the explicit formula here.	In \cref{fig:dNdz_EarlyTimes_MidTimes} we present plots of \cref{eq:MattDistzt,eq:dNdzRad} at two different times, scaled by $ Q_s $ and $ R $ so as to consider dimensionless variables.

%
%

			\begin{figure*}
				\centering
				\begin{subfigure}[b]{0.5\linewidth}
					\includegraphics[scale=0.45]{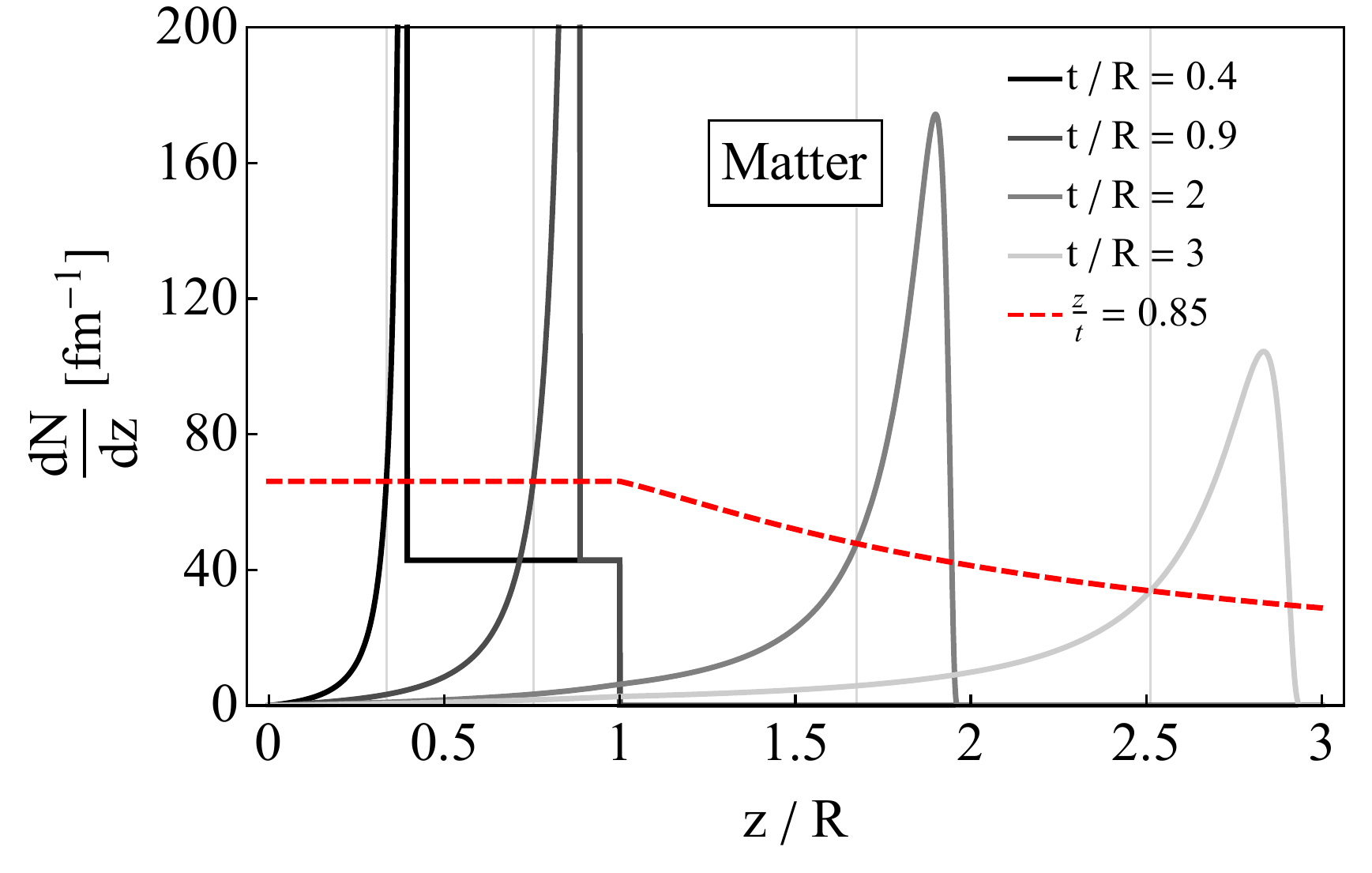}
					\caption{\label{fig:dNdz_ConstZoverT}}
				\end{subfigure}
				\begin{subfigure}[b]{0.5\linewidth}
					\includegraphics[scale=0.45]{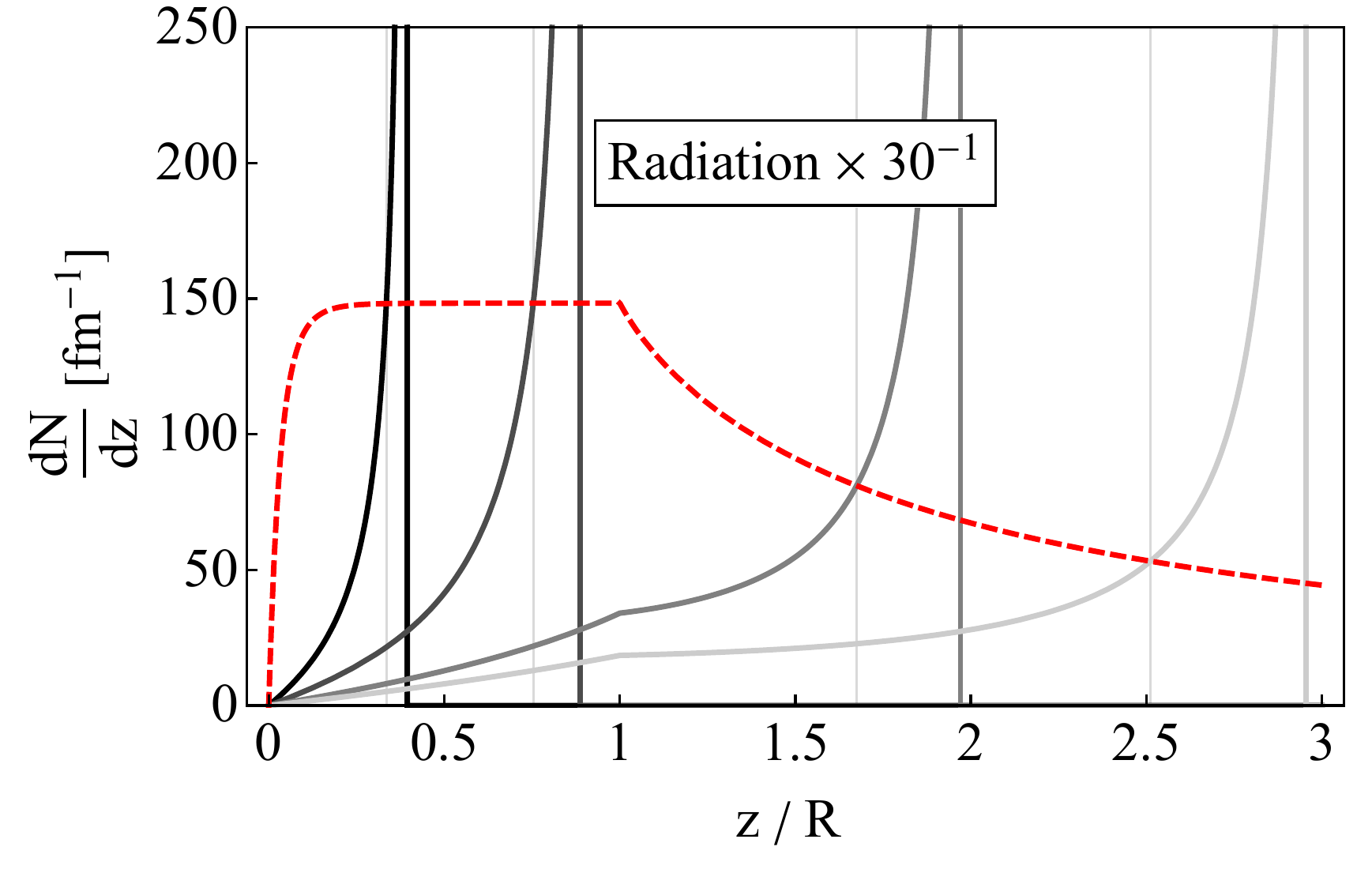}
					\caption{\label{fig:dNdz_ConstZoverT_radiation}}
				\end{subfigure}
				\caption{$ \frac{dN}{dz} $ as a function of time for (a) the matter distribution and (b) the radiation (with formation time) distribution.  The red curves connect points on the (continuous set in $ t $) $ (z,t) $- curves that have $ \sfrac{z}{t}=0.85 $.  The feint vertical lines are at $ z $ values that correspond to $ \frac{z}{t}=0.85 $ for each $ t $-curve.   $ R=14  $ fm/c, $ N=600 $, $ F=30 $, $ Q_s =1 $ GeV.
					\label{fig:dNdz_ConstZoverT_both}}
			\end{figure*}
		
		At early times, the peak in the matter distribution, shown in \cref{fig:dNdz_EarlyTimes}, indicates that the $ (z,t) $-distribution of the matter particles exhibits the expected compression as a result of the collision.  \Cref{fig:dNdz_MidTimes} shows that the matter will continue to expand after the shock-wave has passed  (the probability density falls off). 
		
		Since the problem is framed relativistically, the compression naturally has a component that is simply due to Lorentz contraction.  However, one may boost into a ``co-moving frame''
		\footnote{At early times it is hard to interpret this frame as a genuinely co-moving frame since the fluid has many components all moving at different velocities, it is not clear what the frame is co-moving with.  However, at late times (once the sheet has passed through the nucleus), any frame at rapidity $ \eta_s $ will be moving along with a fluid who's average velocity is $ \langle y\rangle=\eta_s$. }
		(of constant $ \eta_s $ corresponding to a constant $ \sfrac{z}{t} $) where one will find that the density of particles at a given $ \eta_s $ is constant in the spatial region of the initial matter distribution, and falls off beyond $ R $.  To see this effect, we present \cref{fig:dNdz_ConstZoverT}, in which the thin vertical lines are lines of constant $ \frac{z}{t}=0.85 $, and the different solid curves show the probability density at progressively higher $ t $.  Notice in \cref{fig:dNdz_ConstZoverT} that, for the first two solid curves (where $ t < R $), the two points in the same co-moving frame (the intersection of the vertical lines with the colored curves) are at the same height, while the points in this same co-moving frame at later times are at lower probability density.  The probability density for constant $ \frac{z}{t}=0.85 $ are shown as a function of $ z $ in the dashed red curve.  The absolute density depends on the choice of co-moving frame, with frames closer to the frame of the shockwave ($ \sfrac{z}{t} \rightarrow 1$) presenting higher densities than those frames that lag behind.  Although the radiation curve in \cref{fig:dNdz_ConstZoverT_radiation} also exhibits a fall off in density at a fixed $ \eta_s $ beyond $ z=R $, the density in the co-moving frame climbs throughout $ z<R $, since new radiation is added successively as the sheet passes through the matter.

		 The divergence observed in the radiation is an artifact of the flat production distribution, meaning that a large number of particles are radiated at relativistic speeds.  This effect is apparent already in \cref{eq:dNdzRad2} where a divergence occurs as $ z\rightarrow t $.  Alternatively: the divergence is an artifact of the choice of frame and disappears when one considers the radiation in a comoving frame.  Consider, for instance, the frame with $ \eta_s = 1.25 $ (corresponding to $ \sfrac{z}{t}=0.85 $), plotted as a dashed red curve in \cref{fig:dNdz_ConstZoverT_radiation}, showing that the number density of the radiation in a comoving frame will build up while the frame is moving through the target nucleus, before dropping off (due to expansion) beyond the nucleus.  We have not plotted the radiation with formation time here as the formation does not have an observable effect in $ (z,t) $-space.  The matter distribution does not suffer from the same effect since we have used a more realistic distribution of the momenta imparted to the matter particles.

	 	Both the matter and the radiation curves in \cref{fig:dNdz_MidTimes} exhibit a ``kink'' at $ z=R $.  We may understand the kink as resulting in the following way;  since the shockwave imparts a distribution of velocities to the matter particles situated at $ z\leq R $, some pile-up may occur as particles accorded higher velocities at earlier times catch up with those particles that obtained lower velocities at later times.  However, once the shockwave passes the limits of the nucleus, no further pile-up will occur.  The particles will then successively leave the region of initial distribution as time passes, while particles with higher velocities will move off more rapidly (causing a lower particle density beyond the range of the initial matter distribution). 
	 	
	 	If one includes a formation time for the radiation of the form		
		 	\begin{equation}
		 	N_0\left(1-\e^{-\tau/\tau_0}\right)\overset{\tau\gg\tau_0}{\longrightarrow}N_0,
		 	\end{equation}
		 and a limit to the beam rapidity of $ y_b $, one can follow the same process and arrive at a result for the number density of particles radiated with a formation time of $ \tau_f \sim \sfrac{1}{Q_s}$ by at-rest quarks that are struck by a relativistic sheet of colored glass with rapidity $ y_b $.  The full result, derived in detail in \cref{app:RadDerivztTauform}, resembles \cref{eq:dNdzRad} in form but requires a slightly more involved integral:
		 	\begin{align}\label{eq:dNdzRad3}
		 	I^{\rad}_{\text{N},\tau_f}(a)
		 	&=\int_{\tanh^{-1}(a)}^{\tanh^{-1}(\sfrac{z}{t})}dy\,
		 	\frac{v_b}{v_b-\tanh y}\left(1-\exp\left\{-\frac{Q_s(z-tv_b)}{\tanh y-v_b}\sech y\right\}\right).
		 	\end{align}

\begin{figure*}
	 			\centering
	 			\includegraphics[scale=0.6,trim=0 2cm 0 1.cm]{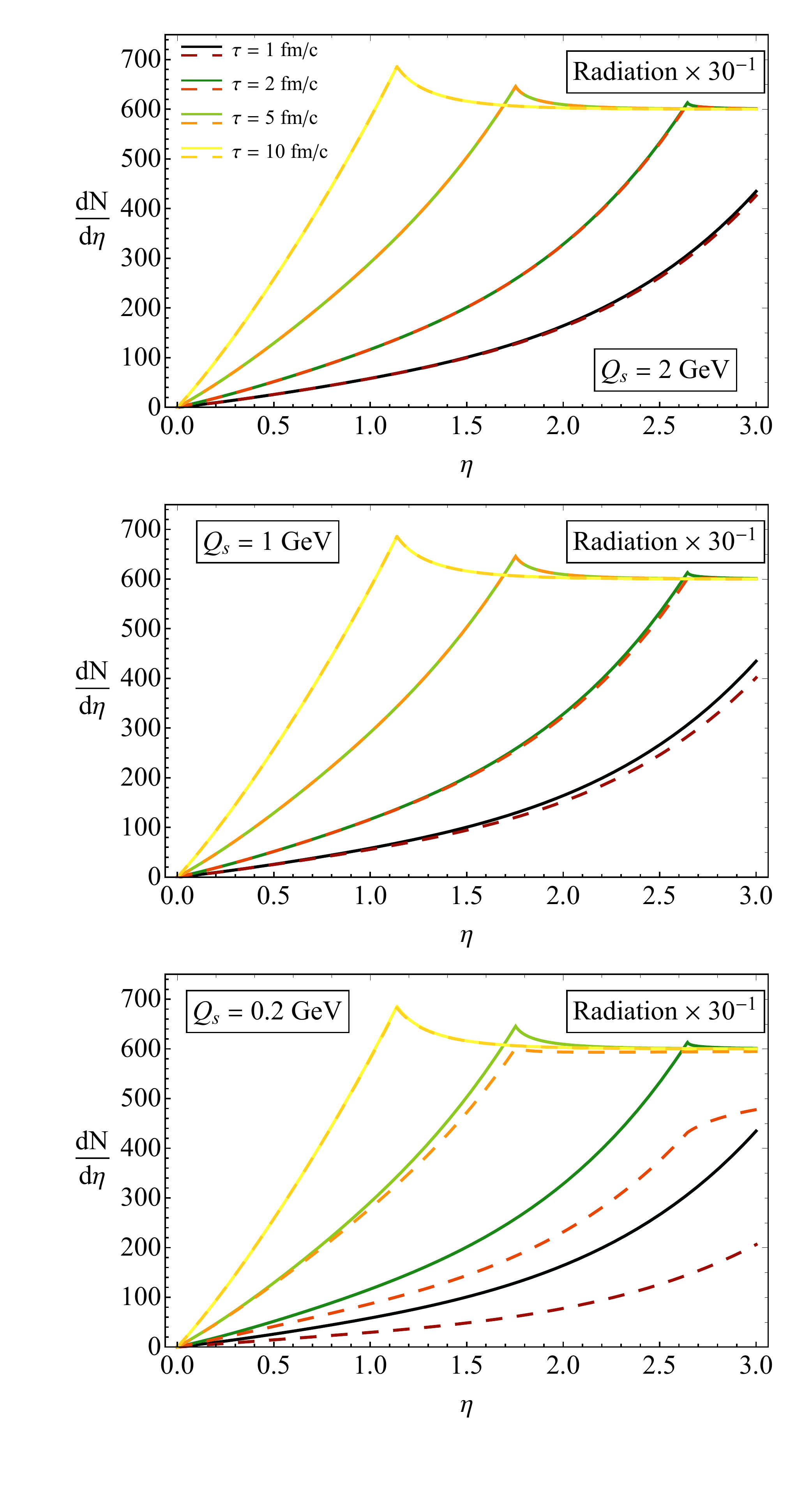}
	 			\caption{The radiation profile at various $ \tau $ without formation time (solid curves) and with formation time (dashed curves), described by appropriately transformed \cref{eq:dNdzRad}.  Progressively lighter curves are at later times. $ R=14  $ fm/c, $ N=600 $, $ F=30 $.	\label{fig:dNdEta_ManyTimes_Radiation_all} }
	 			
	 		\end{figure*}

	 	The integral in \cref{eq:dNdzRad3} has a closed form but it is not illuminating.  In \cref{fig:dNdEta_ManyTimes_Radiation_all} we show the effect of the formation time, which we have taken to scale with the inverse of the saturation scale $\tau_f\sim Q_s^{-1} $. The solid green curves show the radiation density without formation time while the orange dashed curves show the radiation with formation time.  Each panel has a different value of $ Q_s $ for the formation time with decreasing $ Q_s $ (increasing $ \tau_f $) downwards.  The solid green curves are then identical in all three plots.  Each panel shows four curves, each representing the density at a different time with lighter curves (further to the left of each plot) at later times.  One notices immediately that at late times (light curves), the formation time plays no role - the dashed curves lie on top of the solid curves. We also see that a larger formation time (panels lower down) means that the dashed curves take longer to resemble the solid curves.  Equivalently, larger formation times restrict the production of radiation for longer, as expected.

	\subsection{Energy density}\label{sec:dEdEta}
		The energy density associated with a distribution of particles can be extracted from the phase-space density $f(x,p)$ in the following way:
		\eqn{\frac{dE(x)}{dz}&=\int dp f(x,p)E(p)\label{matt_energy_density}\,,
		}
		where $E(p)$ is the energy of a particle carrying a $z$-component of momentum $p^z=p$. The matter and the radiation have different dispersion relations. 
		
		Note that the full energy should be relativistically divergent in the lab- or rest-frame of the target nucleus.  The full energy should diverge as $ Q_s \cosh\eta_s $.  One may boost into a frame which has a particular rapidity $ \eta_s $ in the lab frame, but it is difficult to interpret this frame as a ``comoving'' frame at early times, since it is only at late times that all the particles at a particular $ \eta_s $ are moving with the same $ y $ such that $ \langle y\rangle=\eta_s $.  We will therefore call such a frame an $ \eta_s $-moving frame.  Boosting to an $ \eta_s $-moving frame simply involves dividing by the Lorentz-$ \gamma $ factor, $ \gamma=\cosh\eta_s $.  In this way then, we will not do a full Lorentz boost of the current-density four-vector. The transverse energy is  Lorentz-invariant.

			In the case of the matter distribution the energy is given by
			\eqn{E(p)&=\sqrt{p^2+p_{\text{\tiny{T}}}^2+m^2}=p+m
			} 
			and the transverse energy is
			\eqn{E_T(p)&=\sqrt{p_T^2+m^2}\nonumber\\
				&=m\sqrt{1+\frac{2p}{m}}\,,
			}
			where we have made use of the following relationship between the transverse and longitudinal momentum of the struck matter particle \footnote{Also given in Eq.\eqref{pT-pz}}
			\eqn{p_T^2=2mp\,\,.
			} 
		
			Using Eq.\eqref{matt_energy_density} we arrive at the following result for energy density of the matter distribution
				\begin{align}
					\frac{dE^{\mat}(x)}{dz}
						&=\int dp f(x,p)(p+m) \nonumber\\
						&=\frac{N}{R}\theta(z-t)\theta(z)\theta(R-z)m+\frac{N}{R}\theta(t-z)\left\lbrace I_{E}^{\mat}\left(\frac{z}{t}\right)-\theta(z-R)I_E^{\mat}\left(\frac{z-R}{t-R}\right) \right\rbrace\,,
						\label{eq:dEdzMatt}
				\end{align} 
			where 	
			\eqn{I_E^{\mat}(w)&=\int_{0}^{\frac{w}{1-w}m}dp\frac{(p+m)^2}{mp_0}e^{-\frac{p}{p_0}}\\
				&= m+2p_0+\frac{2p_0^2}{m}-\left(\frac{m}{(1-w)^2}+\frac{2p_0}{1-w}+\frac{2p_0^2}{m} \right)\exp\left(-\frac{m}{p_0}\frac{w}{1-w}\right)\, .
			}
			The transverse energy has a similar formula
			\eqn{\frac{dE_T^{\mat}(x)}{dz}&=\int dp f(x,p)\sqrt{2mp+m^2} \nonumber\\
				&=\frac{N}{R}\theta(z-t)\theta(z)\theta(R-z)m+\frac{N}{R}\theta(t-z)\left\lbrace I_{E_T}^{\mat}\left(\frac{z}{t}\right)-\theta(z-R)I_{E_T}^{\mat}\left(\frac{z-R}{t-R}\right) \right\rbrace\,,
			}
			where
			\eqn{I_{E_T}^{\mat}(w)&=\int_{0}^{\frac{w}{1-w}m}dp\frac{1}{p_0}(p+m)\sqrt{1+\frac{2p}{m}}e^{-\frac{p}{p_0}}
			}

			In the case of the radiation, we will assume that the transverse momentum is entirely as a result of the momentum transfer from the sheet of colored glass and is therefore on the order of $ Q_s $ so that $ k_{\perp}\sim Q_s $.  Since $ y=\half\ln\frac{k^+}{k^-} $, we have that
				\begin{equation}
					E(y) = \sqrt{k_{\perp}^2+k_z^2} = Q_s\cosh y = Q_s \gamma.
				\end{equation}
			
			The energy density of the radiation (without formation time or beam rapidity limit) is therefore given by
				\begin{equation}
				\frac{dE^{\rad}}{dz}
				=\frac{Q_s}{R}F\,N\,\T{t-z}\Bigg[\T{R-z}I_{E}^{\rad}(0)
				+\T{R-z}I_{E}^{\rad}\left(\frac{z-R}{t-R}\right)\Bigg],
				\label{eq:dEdzRad}
				\end{equation}
			where
				\begin{align}
				I^{\rad}_{E}(a)
				&=\int_{\tanh^{-1} a}^{\tanh^{-1} \zot}dy\frac{\cosh y}{1-\tanh y}.
				\end{align}	
			
			In an $ \eta_s $-moving frame, the energy density is given by
				\begin{align}\label{eq:dEdEta_matter}
				\frac{1}{\cosh\eta}\frac{dE^{\text{rad}}}{d\eta}
				=\frac{1}{\cosh\eta}\frac{\tau}{\cosh\eta}\frac{dE^{\rad}}{dz}\big(z(\eta,\tau),t(\eta,\tau)\big).
				\end{align}
			
			The transverse energy for the radiation is simply given by
				\begin{equation}
				\frac{dE_{T}^{\rad}}{dz} = Q_s\frac{dN^{\rad}}{dz},
				\end{equation}

			with $ \sfrac{dN^{\rad}}{dz} $	given by \cref{eq:dNdzRad2,eq:dNdzRad2}.

	\subsection{Late-time distributions}
	
			\begin{figure}
				\centering
				\begin{subfigure}[b]{0.48\linewidth}
				\includegraphics[scale=0.45]{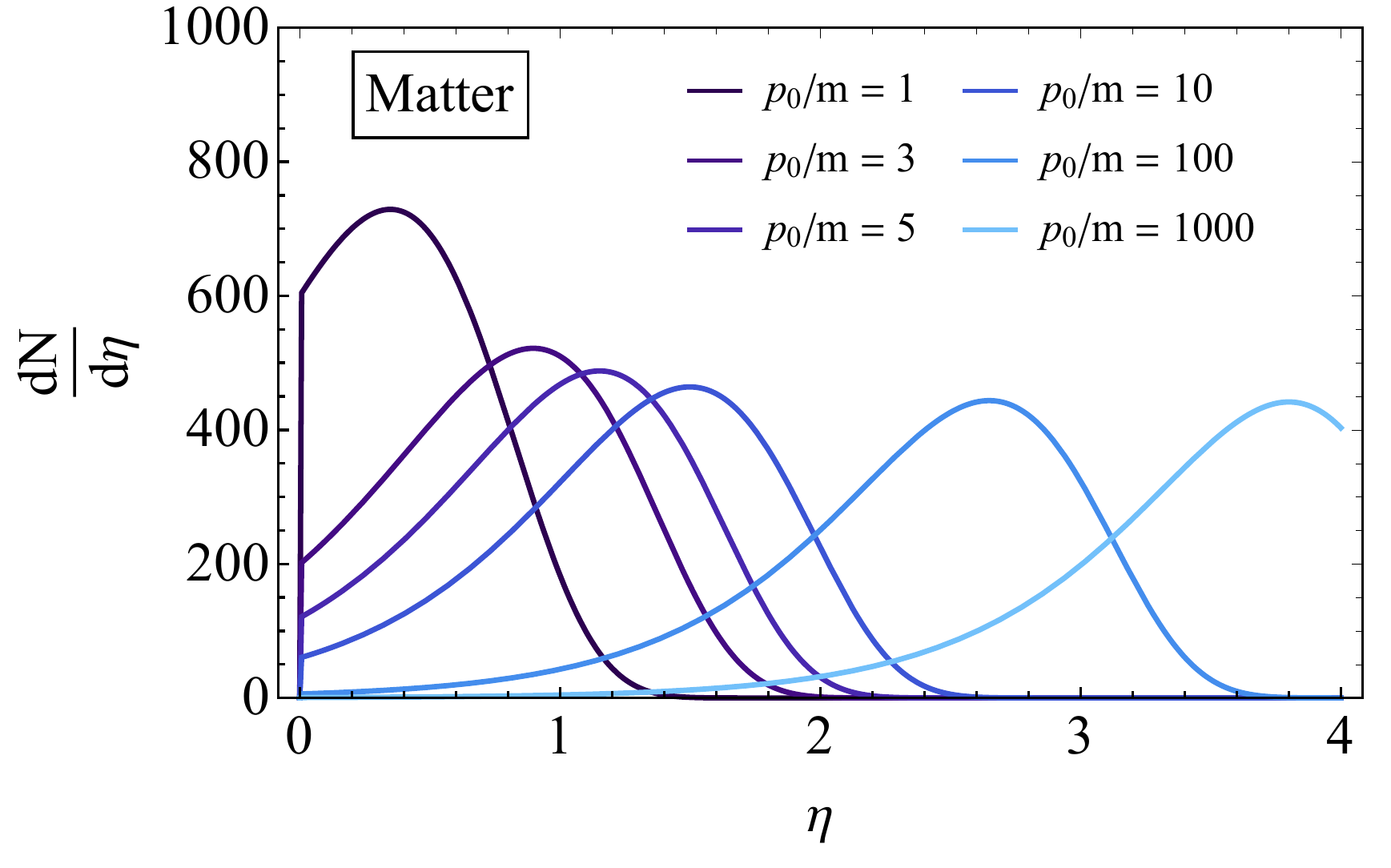}
					\caption{\label{fig:LateTimes_Qs}}
				\end{subfigure}
				\begin{subfigure}[b]{0.48\linewidth}
					\includegraphics[scale=0.45]{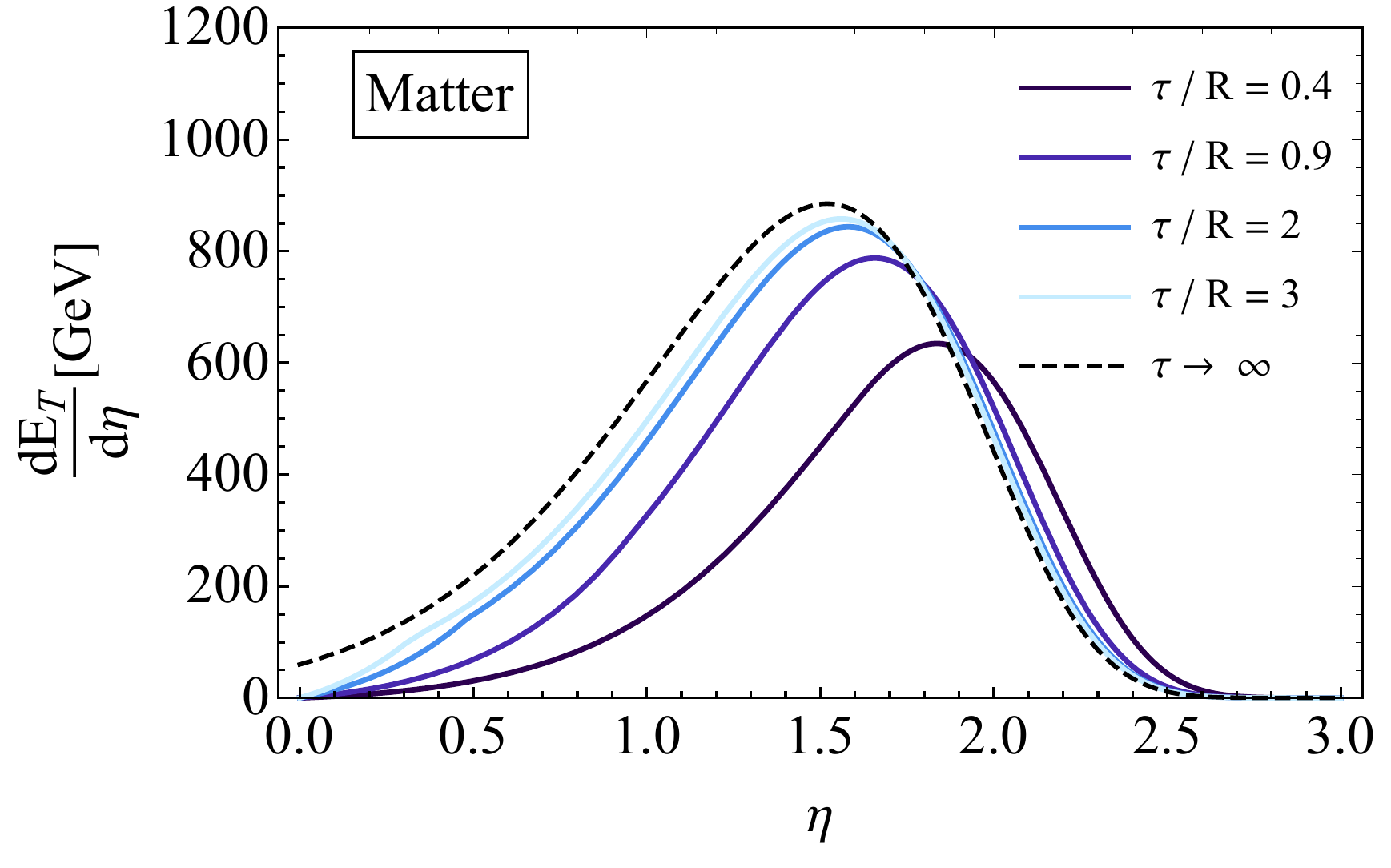}
					\caption{\label{fig:LateTimes_dEdEta}}
				\end{subfigure}
				\caption{The late time distribution of the (a) number density and (b) transverse energy density for the matter, using $ R=14 $ fm, $ N=600 $  matter particles, and $ \frac{p_0}{m}= 10$. \label{fig:LateTimes_both}}
			\end{figure}	
	
		At late times, $\tau\gg R>0$, the region for which the $\tau \sinh\eta-R<0$ is ever shrinking and the increasingly important region, in $\eta$-space, is that for which   $\eta>\sinh^{-1}(R/\tau)$.  We therefore have that \eqref{matt-dist}, describing the distribution of matter, becomes:
		
		\eq{\frac{dN^{\mat}}{d\eta}&\rightarrow \frac{m}{p_0}N\exp\left\lbrace 2\eta-\left(\frac{m}{2p_0}\right)e^{2\eta} +\left(\frac{m}{2p_0}\right)\right\rbrace\,\,. \this\label{matt-limit}
		}
		The late-time matter distribution for the matter \eqref{matt-limit} has a peak at $\eta=\frac{1}{2}\log\frac{2p_0}{m}$, with a height given by
		\eqn{\left.\frac{dN^{\mat}}{d\eta}\right|_{\text{max}}=2Ne^{\left(\frac{m}{2p_0}-1\right)}\,\,.\label{eq:LateTimeMatterPeak}
		}
		
		This means that baryons lose a rapidity of $\log\frac{Q_s}{m}$, which is $1-2$ units of rapidity at RHIC energies, and $2-3$ units of rapidity at the LHC. It is interesting to see that this is in agreement with the results of \cite{Busza:1983rj,Busza:2018rrf}, where very different methods are used. In figure \cref{fig:LateTimes_Qs} we present an investigation of the influence of $ \sfrac{p_0}{m} $ on the position and height of the peak in \cref{matt-limit}.	
		
		We may also consider the late time distribution of the radiation, which is almost trivial since we started out with the assumption that the momentum-space rapidity distribution of the radiation is flat.  Since the momentum-space rapidity is equal to the position-space rapidity at late times, we therefore expect the late-time position-space rapidity of the radiation to be flat as well.  In this limit, the radiation distribution becomes (after expansion around $ \tau\rightarrow\infty $)
			\begin{align}
				\frac{dN^{\rad}}{d\eta}\Bigg\vert_{\tau\rightarrow\infty}
				&=\lim_{\tau\rightarrow\infty}\left\{
				\frac{R}{\tau}\e^{\eta}-\half\ln\left[1-\frac{2R}{\tau}\e^{-\eta}\right]\right\}
				\frac{\tau}{\cosh\eta}\frac{NF}{2R}
				=NF.
			\end{align}
			
		At late times then, the radiation is constant in space-time rapidity, $ \eta_s $, provided their production is flat in momentum rapidity $ y $.

		In an exactly analagous manner, one may study the late time distribution of the energy density.  For the radiation this is a trivial exercise of multiplying the number density by $ Q_s $ - each radiated particle caries a transverse energy of around $ Q_s $.  A far more interesting exercise is to consider the late time distribution of the matter.  The same process of expanding \cref{eq:dEdEta_matter} around $ \tau\rightarrow\infty $ yields the following late time expression
			\begin{align}
				\frac{dE_T^{\mat}}{d\eta}\rightarrow 
				\frac{N m^2}{p_0}\frac{\sech^3\eta}{(1-\tanh\eta)^3}\exp\left[-\frac{m}{p_0}\frac{\tanh\eta}{1-\tanh\eta}\right],
			\end{align}
		which has a peak at $ \eta=\log\sqrt{\frac{3p_0}{m}} $ with height
			\begin{equation}\label{eq:LateTimeMatterPeak_Energy}
				\left.\frac{dE_T^{\mat}}{d\eta}\right\vert_{max}=
				3\sqrt{3 p_0 m}\,N  \exp\left[-\half\left(3-\frac{m}{p_0}\right)\right].
			\end{equation}

		The late-time distribution of the energy density for the matter is shown in \cref{fig:LateTimes_dEdEta}.   Notice that the distribution of matter is peaked in $ (\eta,\tau) $-space at late times, the peak being described by \cref{eq:LateTimeMatterPeak}.  This peak is characteristic of the distribution and is related to saturation physics through $ \sfrac{p_0}{m} \sim\sfrac{Q_s^2}{m^2}$.  The position of the peak moves along $ \eta $ as $ \ln \sfrac{Q_s^2}{m^2} $ while the height of the peak scales with $ \exp\left(\sfrac{m^2}{Q_s^2}\right) $, which may be seen from the relations in \cref{app:Deriv_matter} and the fact that $ \mu^2\sim Q_s^2 $.  The dependence on $ \sfrac{p_0}{m} $ is plotted in \cref{fig:LateTimes_Qs}.
		
		One may note that the position of the peak for the late time number density distribution in \cref{eq:LateTimeMatterPeak} is different from the position of the peak for the late time energy density in \cref{eq:LateTimeMatterPeak_Energy}.  The difference is only by a constant, dimensionless number which is small ($ \sim 3 $) in comparison with typical values of $ \sfrac{p_0}{m}\gtrsim 10 $.

\section{Average velocity and momentum-space rapidity}\label{sec:AvgVandY}
	
	Information about the local velocity of the quarks should be contained in the quark phase-space density $f(x,p)$ given by Eq. \eqref{Matt-dist}. We extract it by first constructing a probability density in $p$ for a given $z,t$.
	Now, the total number of particles at $(z,t)$ is given by:
	\eqn{dN(z,t)&=\left(\int_{0}^{\infty}dpf(x,p)\right) dz.
	}   
	The number of particles at $(z,t)$ with momentum $p'$ is $f(x,p')\,dzdp'$, so the probability distribution is 
	\eqn{d\text{Pr}(p'|z,t)&=\frac{f(x,p')dz dp'}{\left(\int_{0}^{\infty}dpf(x,p)\right)dz} \nonumber \\
		\implies \frac{d\text{Pr}(p'|z,t)}{dp'}&= \frac{f(x,p')}{\int_{0}^{\infty}dpf(x,p)} \,\,.\label{prob-density-matt-def}
	}
	This probability density can be written explicitly as
	\eqn{\frac{d\text{Pr}(p|z,t)}{dp}&=\theta(t-z)\left\lbrace\frac{\theta(R-z)\theta\left(\frac{zm}{t-z}-p \right)\theta(p)}{I_0^\mat\left(\frac{z}{t}\right)}\frac{p+m}{mp_0}e^{-\frac{p}{p_{0}}} +\frac{\theta(z-R)\theta\left(\frac{zm}{t-z}-p \right)\theta\left(p-\frac{z-R}{t-z}m\right)}{I_0^\mat\left(\frac{z}{t}\right)-I_0^\mat\left(\frac{z-R}{t-R}\right)} \frac{p+m}{mp_0}e^{-\frac{p}{p_{0}}}\right\rbrace \nonumber\\
	&+	\theta(z-t)\frac{\theta(z)\theta(R-z)m}{(p+m)^2}\delta\left(\frac{p}{p+m}\right).	
	}
	We calculate the averages of momentum and velocity over the above probability density, getting
	\eqn{\langle y\rangle^{\mat} (z,t)&=\int dp\frac{1}{2} \ln\left(\frac{p^+}{p^-}\right) \frac{d\text{Pr}(p|z,t)}{dp}   \nonumber\\
		&= \theta(t-z)\left\lbrace\theta(R-z)\frac{I_y^{\mat}\left(\frac{z}{t}\right)}{I_0^{\mat}\left(\frac{z}{t}\right)} +\theta(z-R)\frac{I_y^{\mat}\left(\frac{z}{t}\right)-I_y^{\mat}\left(\frac{z-R}{t-R}\right)}{I_0^{\mat}\left(\frac{z}{t}\right)-I_0^{\mat}\left(\frac{z-R}{t-R}\right)} \right\rbrace  \,\,,\label{avgYmatt}
	}
	\eqn{\langle v\rangle^{\mat} (z,t)&= \int dp \frac{p^z}{p^0}  \frac{d\text{Pr}(p|z,t)}{dp}   \nonumber\\
		&=  \theta(t-z)\left\lbrace\theta(R-z)\frac{I_v^{\mat}\left(\frac{z}{t}\right)}{I_0^{\mat}\left(\frac{z}{t}\right)} +\,\theta(z-R)\frac{I_v^{\mat}\left(\frac{z}{t}\right)-I_v^{\mat}\left(\frac{z-R}{t-R}\right)}{I_0^{\mat}\left(\frac{z}{t}\right)-I_0^{\mat}\left(\frac{z-R}{t-R}\right)} \right\rbrace\,\,,\label{avgVmatt}
	}

	where we have made use of the integrals
\eqn{I_y^{\mat}(w)&=\int_0^{\frac{w}{1-w}m}\!\!\!dp\frac{p+m}{mp_0}\frac{1}{2}\ln\left(\frac{2p+m}{m} \right)e^{-\frac{p}{p_0}}\,,\nonumber\\
	&=\frac{1}{2}\left\lbrace -\exp\left(-\frac{m}{p_0}\frac{w}{1-w}\right)\ln \frac{1+w}{1-w} + e^{\frac{m}{2p_0}}\textbf{Ei}\left(\frac{m}{2p_0}\frac{1+w}{1-w}\right) -e^{\frac{m}{2p_0}}\textbf{Ei}\left(\frac{m}{2p_0}\right)\right\rbrace,\label{IyMat}\\
I_v^{\mat}(w)&=\int_0^{\frac{w}{1-w}m}dp\frac{p}{mp_0}e^{-\frac{p}{p_0}} \quad ,\nonumber\\
	&=\frac{p_0}{m}-\left(\frac{p_0}{m}+\frac{w}{1-w}\right)\exp\left(-\frac{m}{p_0} \frac{w}{1-w}\right)\,\label{IvMat}\\
I_0^{\mat}(w)&=1+\frac{p_0}{m}-\left(\frac{w}{1-w}+1+\frac{p_0}{m} \right)\exp\left(-\frac{m}{p_0}\frac{w}{1-w} \right)\label{I0Mat}\,\,.
}
	
		\begin{figure}
			\centering
			\begin{subfigure}[b]{0.5\linewidth}
				\includegraphics[scale=0.45]{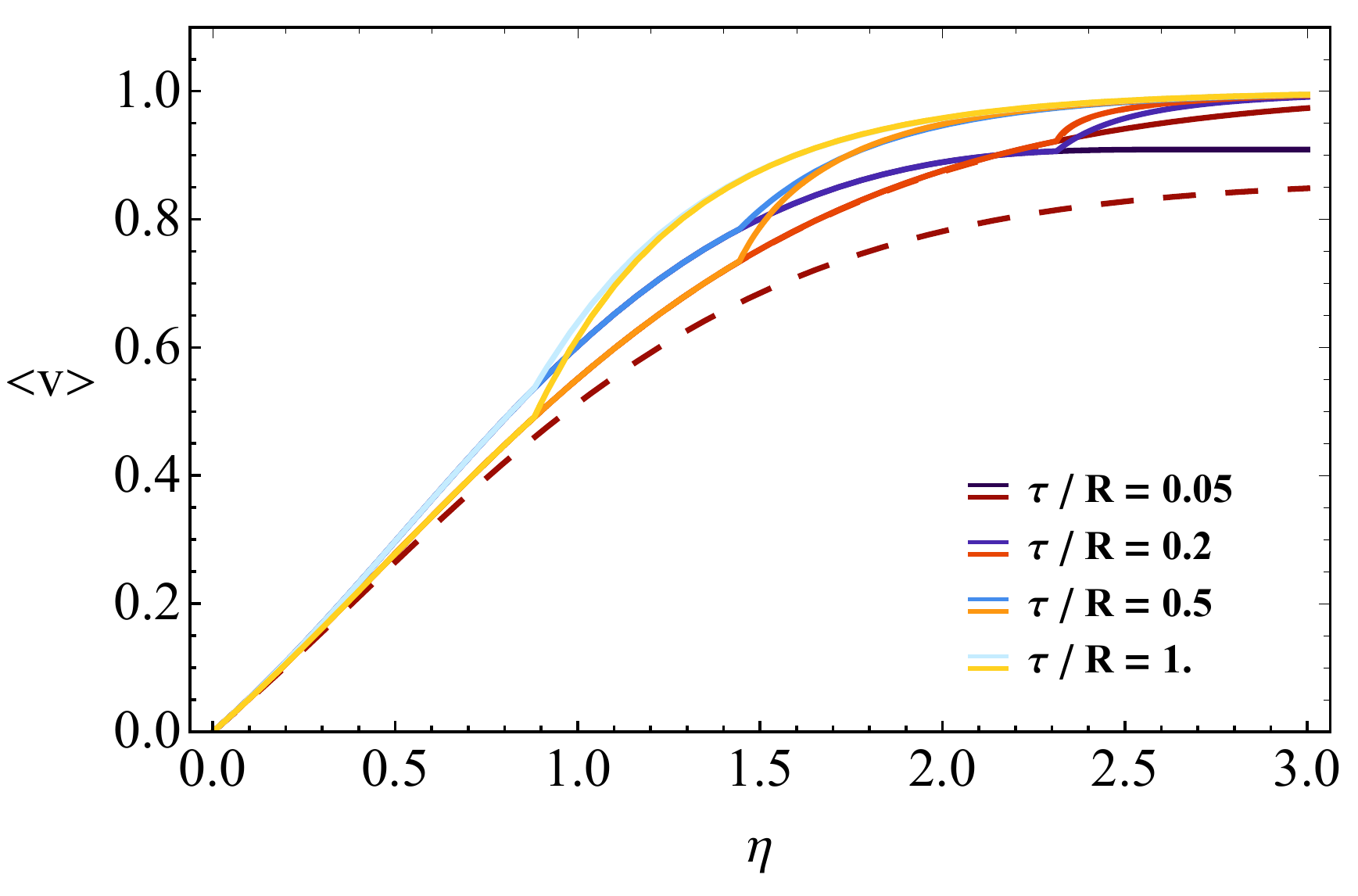}
				\caption{\label{fig:avgVMattRad}}
			\end{subfigure}
			\begin{subfigure}[b]{0.5\linewidth}
				\includegraphics[scale=0.45]{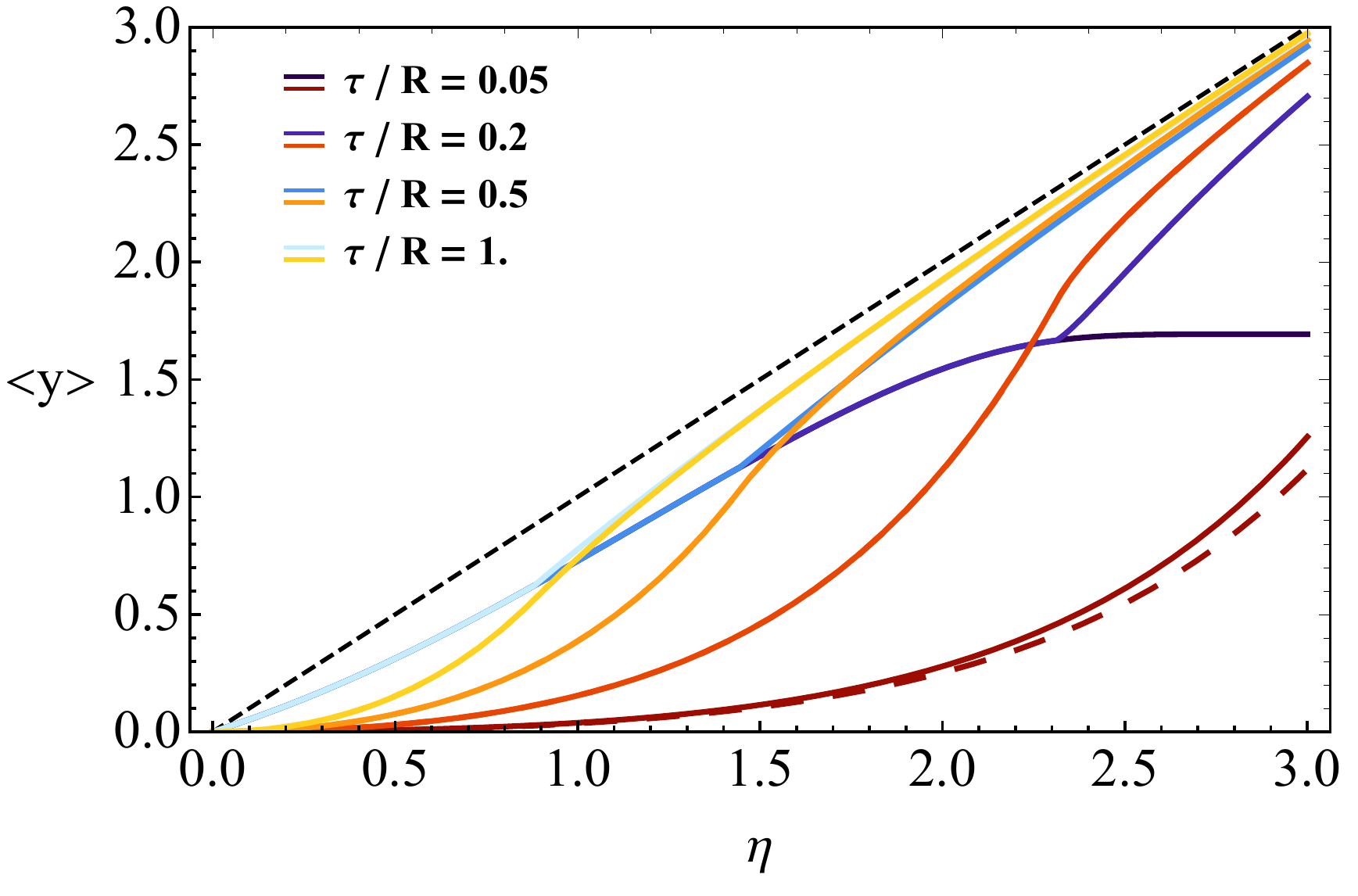}
				\caption{\label{fig:avgYMattRad}}
			\end{subfigure}
			\caption{The average $ v $ and $ y $ for the matter (solid blue curves, given by \cref{avgVmatt,avgYmatt}) and radiation (warm colored curves, see \cref{app:radAvgV}) particles at different proper times as a function of space-time rapidity.  Here again $ \frac{p_0}{m}= 10$, and $ F= $ 30. The solid warm colors are the radiation without formation time while the dashed lines include a formation time of  $\tau_f^{-1}= Q_s =1 $ GeV.\label{fig:avg_V_Y_Matt_Rad}}
		\end{figure}
	
		The average velocity of the radiated particles may be derived by considering where and when a particle needed to have been produced in order to be at a particular position at a later time.  This derivation is performed in \cref{app:radAvgV}. The derivation involves simply counting all the particles that were produced with velocities that allowed them to be at a given $ (z,t) $, giving the following result:			
			\begin{align}
			\langle v \rangle^{\rad}
			=&\Theta(t v_b-z)\Bigg[\frac{1}{z}\Theta(R-z)I_{v}^{\rad}(z)+\frac{1}{R}\Theta(z-R)I_{v}^{\rad}(R)\Bigg],
			\end{align}			
			where
			\begin{align}
			I_{v}^{\rad}(a)
			&=\int_{0}^{a}dz'\Theta(z-z')\Theta(t v_b-z')\frac{z-z'}{t-\sfrac{z'}{v_b}}.
			\end{align}  
			
		A similar formula exists for the average momentum-space rapidity, also given in \cref{app:radAvgV}.	The average velocity and average momentum-space rapidity of the radiation are shown in \cref{fig:avg_V_Y_Matt_Rad}.  A striking feature is that, although the average velocity of the matter is not very different from that of the radiation, the average rapidity of the matter is dramatically different from that of the radiation.  This difference in average rapidity holds even without the inclusion of a formation time and even at relatively late times.

\section{Summary and discussion}\label{sec:SummaryAndDiscussion}
	
	In this paper we have considered the collision of a sheet of colored glass and a nucleus at rest.  We have performed a simple classical calculation, considering a relativistic sheet incident upon stationary matter particles in one dimension, as well as a simple realisation of the resultant radiation.  Our main analytical results, \cref{eq:MattDistzt,eq:dNdzRad}, give an intuitive picture of the dynamics of both the struck quarks and the resultant radiation.  This paper had two main goals: (1) to reframe the predictions of \cite{Anishetty:1980zp} within the context of the CGC, particularly to consider the compression of the target nucleus, and (2) to understand the early-time dynamics of the fragmentation region of a heavy-ion collision.

	The very simple setup explored in this work allows us to address our first goal directly: how much is the nuclear matter compressed by the shock wave, and how does this compression depend on the saturation scale $Q_s$? To answer this, we consider the local density of the matter in a co-moving frame, where the boost factor is given by $1/\gamma~=~\sqrt{1-\left\langle v\right\rangle_\mat^2(z,t)}\,\,$.  The relevant quantity is then
		\begin{align}
			\frac{d{N'}^{\mat}}{dz'}
				&=\frac{1}{\gamma}\left.\frac{dN}{dz}\right|_{t<R,z<t}\nonumber\\
				&=\frac{N}{R}\sqrt{1-\left( 	\frac{I_v^{\mat}
						\left(\frac{z}{t}\right)}{I_0^{\mat}\left(\frac{z}{t}\right)}\right)^2}\,\,I_0^\mat\left(\frac{z}{t}\right) \nonumber \\ 
				&= 	\frac{N}{R}\sqrt{\left(I_0^{\mat}\left(\frac{z}{t}\right)-I_v^\mat\left(\frac{z}{t}\right)\right)
					\left(I_0^{\mat}\left(\frac{z}{t}\right)+I_v^\mat\left(\frac{z}{t}\right)\right)}\,\,.\label{eq:comovingDensity}
		\end{align}

	It is interesting to note that the co-moving density given by \cref{eq:comovingDensity}  depends only on $z/t$. Since $I_0^\mat\left(\frac{z}{t}\right) -I_v^\mat\left(\frac{z}{t}\right)$ and $I_0^\mat\left(\frac{z}{t}\right)+I_v^\mat\left(\frac{z}{t}\right)$ are increasing and positive for $0<z/t<1$, it follows that the density is highest precisely on the light front $z=t$. From Eqs. \eqref{IvMat} and \eqref{I0Mat}, one can read off the maximum density
	\eqn{\text{max}\left\lbrace \frac{dN^{\mat}}{dz'} \right\rbrace &=\frac{N}{R}\sqrt{1+\frac{2p_0}{m}}\,\,. 
	} 

\begin{figure}
	            \centering
	            \includegraphics[height=5cm]{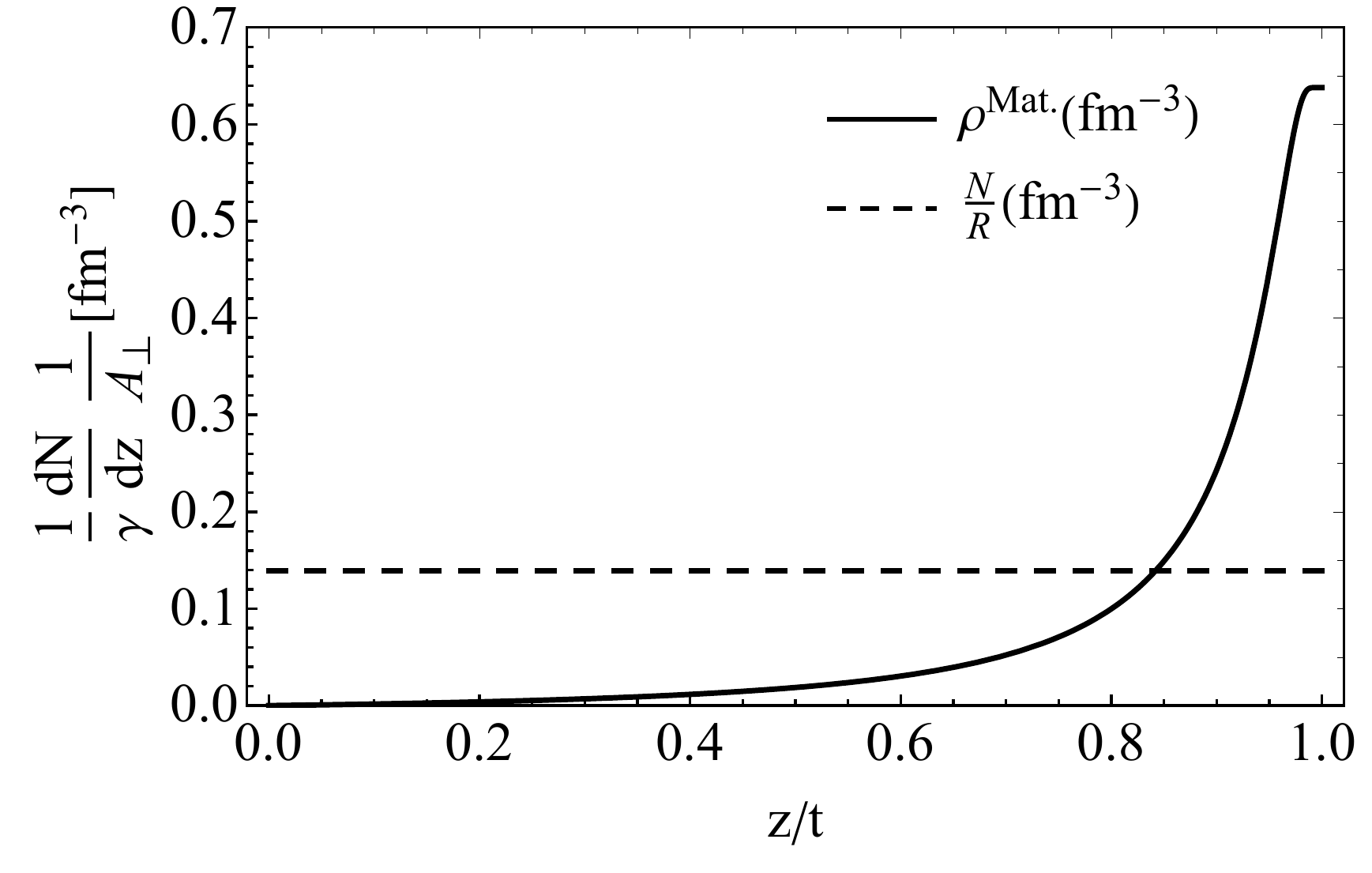}
				\caption{Co-moving density profile of nuclear matter as the color-glass shock wave traverses the nucleus.  The dotted line shows the initial density of matter before the sheet strikes the target nucleus.\label{fig:ComovingDensity}}
	        \end{figure}

	The baryon density is therefore enhanced by a factor of  $\sim \sqrt{p_0/m}\sim Q_s/m$. This means the baryons are compressed by a factor of $5-10$ at RHIC, again it is reassuring that this is in agreement with \cite{Busza:1983rj,Busza:2018rrf}. We have plotted \cref{eq:comovingDensity} in \cref{fig:ComovingDensity}.
	
	        
\begin{figure}
	        \centering
	        \begin{subfigure}[b]{0.5\linewidth}
	        \centering
	        \includegraphics[scale=0.52,trim= 1cm 5cm 1cm 0]{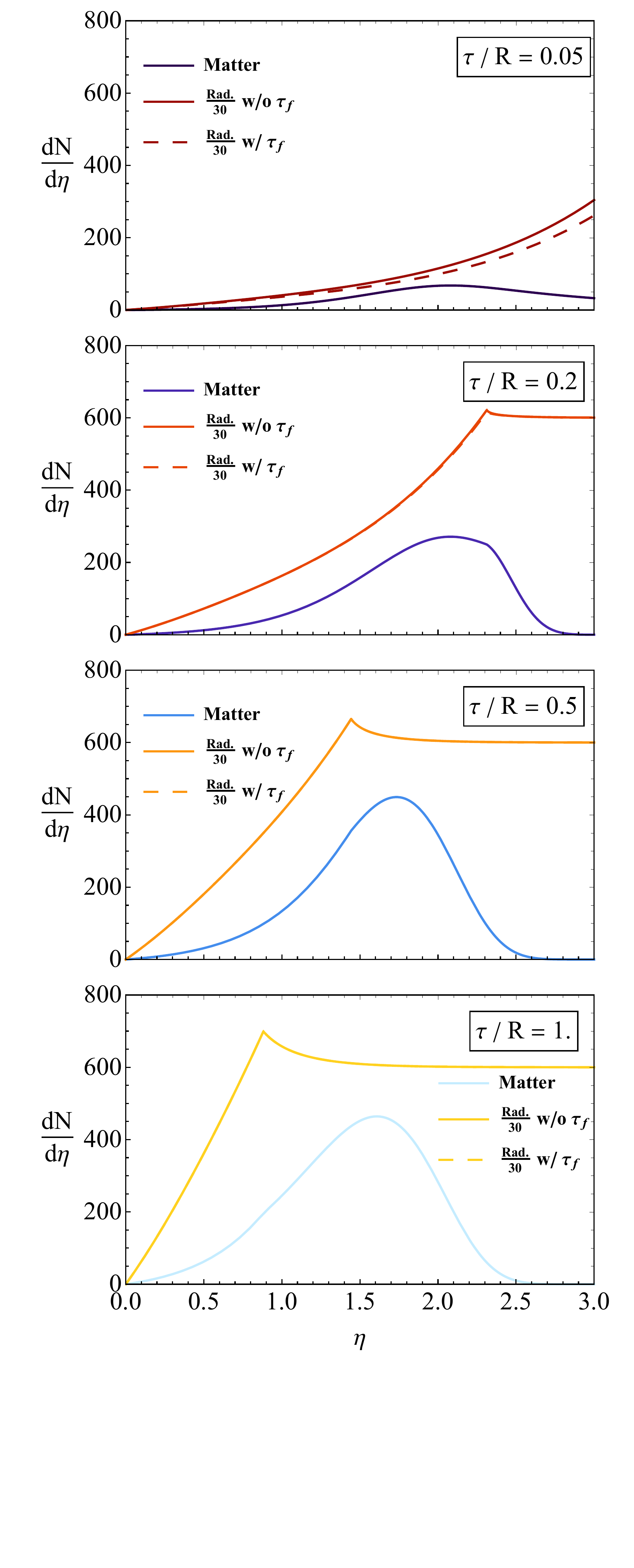}
	        \end{subfigure}\begin{subfigure}[b]{0.5\linewidth}
	        \includegraphics[scale=0.52,trim= 1cm 5cm 1cm 0]{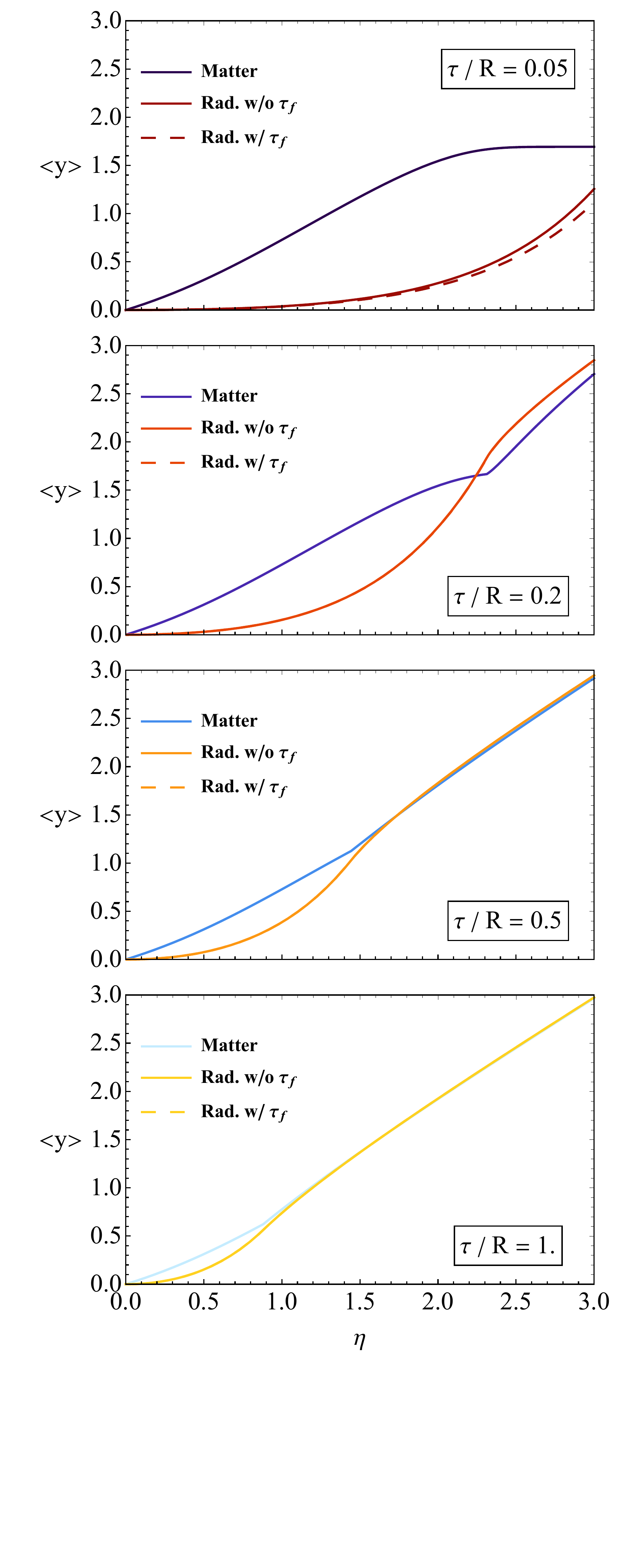}
	        \end{subfigure}
	        \caption{The number density (left column) and the average momentum rapidity (right column) for the matter (solid blue curves), the radiation without formation time (solid orange curves) and the radiation with formation time (dashed orange curves), as a function of the space-time rapidity, for proper time progressing from top to bottom. The radiation curves in the left-hand column have been scaled down by multiplying by $ 30^{-1} $.\label{fig:dNdeta_avgY_all}}
	    \end{figure}

	 \Cref{fig:dNdeta_avgY_all} addresses our second goal by showing the number density (left column) and the average momentum rapidity (right column) for the matter (solid blue curves), the radiation without formation time (solid orange curves) and the radiation with formation time (dashed orange curves), as a function of the space-time rapidity, for proper time progressing from top to bottom.  Notice that, for early times, the average momentum-space rapidity of the matter is very different from that of the radiation.  This is even true for times that are late enough to reduce the effect of the formation time, say $ \tau \sim 3 Q_s^{-1} $.  If we believe that much of the relevant physics is captured in our simple treatment, then the fact that the matter and the radiation have very different rapidity distributions at early times suggests that a full treatment of the early time dynamics in the fragmentation region must necessarily include the dynamics of two fluids.

	We offer a final piece of insight: Although we realize that, in truth, the system we are considering is not yet thermalized, one may gain some intuition of the properties the system might have if it should thermalize, by consider an ideal gas of (two flavor) quarks, anti-quarks, and gluons, described by the thermodynamic pressure \cite{Vogt:2007zz,Cleymans:1985wb}
	    \begin{align}
	        P_{q\bar{q}g} = \frac{37}{90}\pi^2T^4+\mu^2T^2+\frac{1}{2\pi^2}\mu^4.
	    \end{align}    
	
\begin{figure}
	        \centering
	        \begin{subfigure}[b]{0.5\linewidth}
	        \centering
	        \includegraphics[scale=0.5]{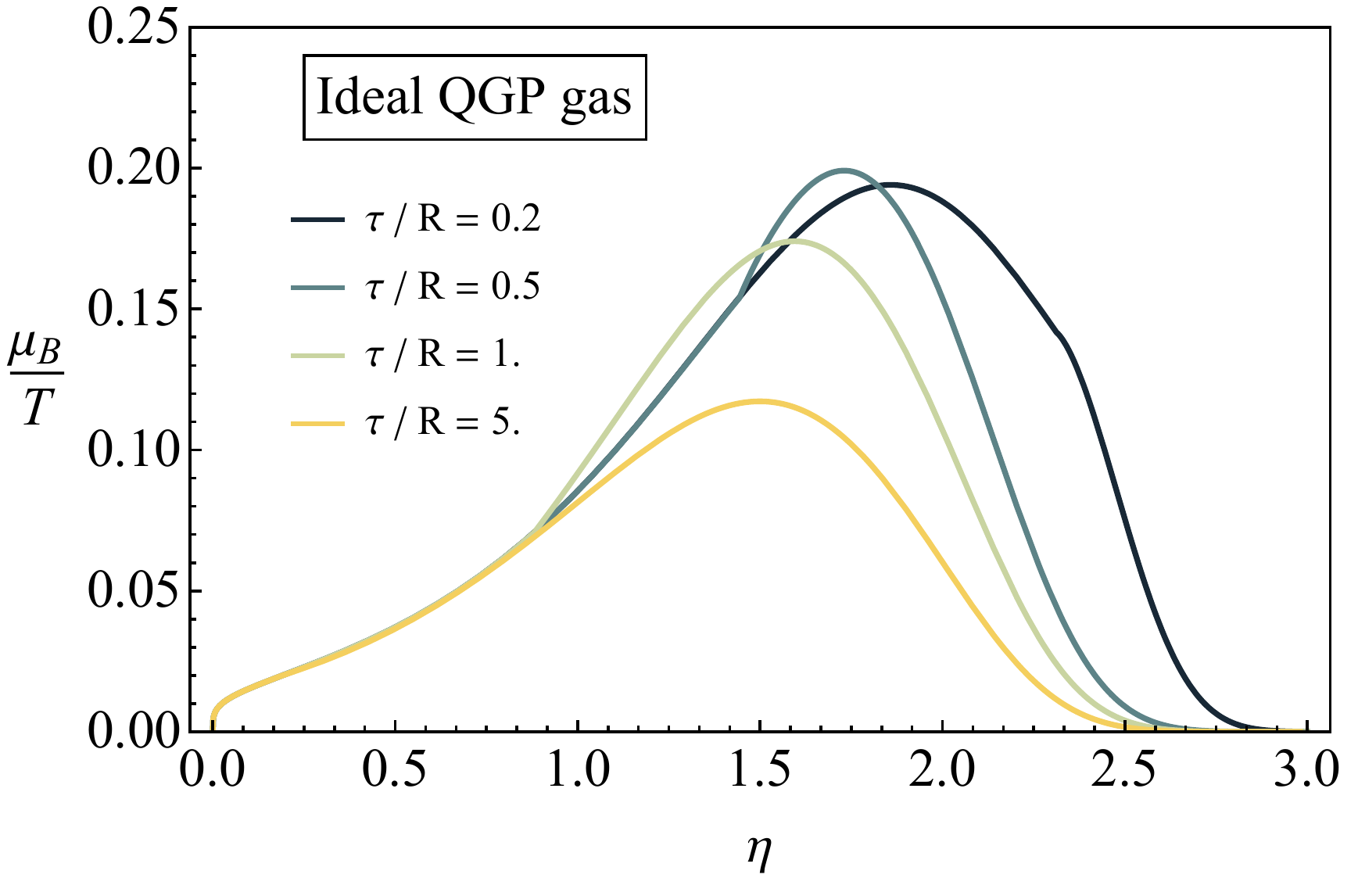}
	        \caption{\label{fig:MuBoT_zot}}
	        \end{subfigure}\begin{subfigure}[b]{0.5\linewidth}
	        \includegraphics[scale=0.5]{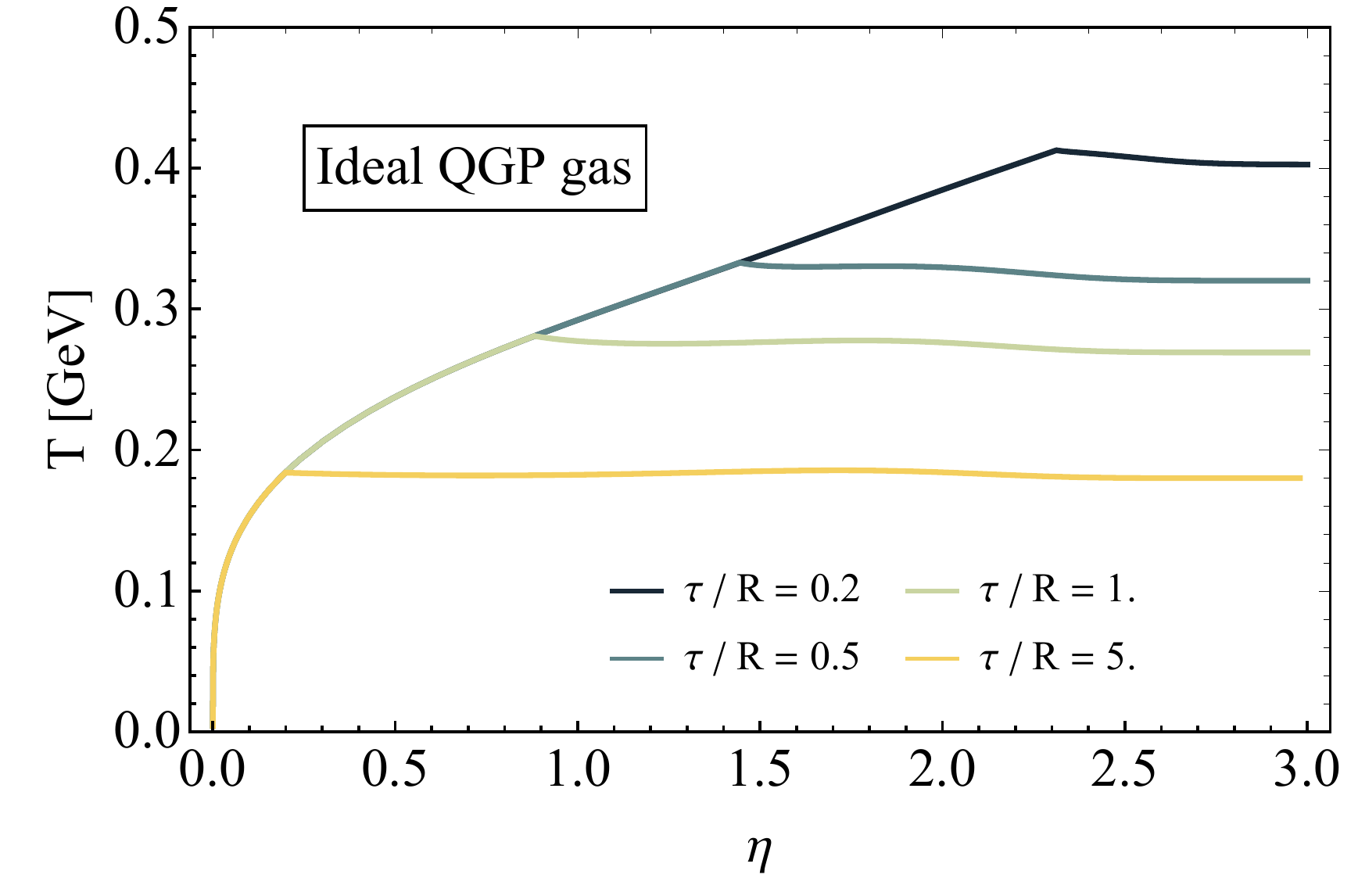}
	        \caption{\label{fig:T_etaTau}}
	        \end{subfigure}
	        \caption{(a) The baryon chemical potential and (b) the temperature, as a function of the space-time rapidity for an ideal gas of quarks, anti-quarks and gluons, at various proper times.}
	        \label{fig:idealGas}
	    \end{figure}

	From the pressure we may compute \cite{Kapusta:2006pm} the number density of the quarks $ n_q=\sfrac{\partial P}{\partial \mu_q} $, the entropy density $ s=\sfrac{\partial P}{\partial T} $, and the energy density $ \epsilon=-P +Ts+n_q\mu_q $.  We can then use the expressions derived in this work for the number density of the matter, along with the sum of the energy density of the radiation and the matter, to  invert the thermodynamical quantities in order to obtain $ \mu_B=3\,\mu_q$, and $ T $. We present, in \cref{fig:idealGas}, the temperature-scaled baryon chemical potential, and the temperature, both as a function of the space-time rapidity for various proper times.  We note that the baryon chemical potential falls off with time in accordance with the $\sfrac{1}{\tau}$ expansion, as does the temperature.  The qualitative behaviour of the chemical potential is naturally dominated by the number density of the matter distribution, so that the shape of the curve in \cref{fig:MuBoT_zot} resembles the shape of the corresponding curve in \cref{fig:LateTimes_Qs}.  Furthermore, the energy density of the fireball is dominated by the energy density of the radiation, so that the qualitative behaviour of \cref{fig:T_etaTau} resembles that of \cref{fig:dNdEta_ManyTimes_Radiation_all}.
	
	We note that, for large enough values of $\eta$ and $\tau$ (such that $\langle y\rangle\sim\eta$ and the system is therefore thermal), one may really consider the curves in \cref{fig:idealGas} to represent the rest-frame of the matter. 
	
\section*{Acknowledgments}
    	
    We wish to gratefully acknowledge many useful discussions with Kiesang Jeong, Keijo Kajantie, Risto Paatelainen, Michal Praszalowicz, Bjoern Schenke, and Wilke van der Schee . Larry McLerran was supported, and  Isobel Kolb\'{e}, Mawande Lushozi,  Gongming Yu were partially supported by
    the U.S. DOE under Grant No. DE-FG02-00ER41132.  Isobel Kolb\'{e}, Mawande Lushozi and Gongming Yu were partially supported under the Multifarious Minds grant provided by the Simons Foundation.  The research of Gongming Yu is supported by the National Natural Science Foundation of China under Grant No. 11847207, the International Postdoctoral Exchange Fellowship Program of China under Grant No. 20180010, and the China Postdoctoral Science Foundation Funded Project under Grant No. 2017M610663.


	\appendix

\section{ Momentum distribution of a classical particle  interacting with a sheet of colored glass}\label{app:pT-dist} 
             It can be shown that, upon interacting with a sheet of color glass, a classical color-charged point particle gets a transverse momentum-kick \cite{Kajantie:2019hft,Kajantie:2019nse}
             \eqn{\frac{\partial p^i}{\partial x^-} &=gT\cdot F^{+i}\,\,,   
             }
            where $F^{\mu\nu}$ is the field associated with the sheet. In $A^+$ gauge, we have:  $F^{+i}= -\partial^{i}A^{+} $, and also
            \eq{A^+&=\frac{1}{\nabla_T^2}\rho(x^-,x_T)=\frac{1}{\nabla_T^2}\rho(\vec{x}_T)\delta(x^-)\,\,\\
            \implies -p^i(x_\perp)&=-gT^a\frac{\partial^{i}}{\nabla_T^2} \rho^a(\vec{x}_\perp)
            }
            The operator $\frac{\partial^i}{\nabla^2_T}$ is defined with the help of the Green's function of the two-dimensional Laplacian:
            \eqn{\left(\frac{\partial^i}{\nabla^2_T}f\right)(\vec{x}_{ _\perp})&=\frac{\partial}{\partial x^i}\int d^2y\frac{1}{2\pi} \ln|\vec{x}_{ _\perp}-\vec{y}_{ _\perp}|f(\vec{y}_{ _\perp})\\
            &= \frac{1}{2\pi}\int d^2y \frac{x^i-y^i}{\left|\vec{x}_{ _\perp}-\vec{y}_{ _\perp}\right|^2}f(\vec{y}_{ _\perp})	\,\,  ,
            }
            and the above integral is only meaningful with the inclusion of an infra-red and ultraviolet cut-off. Setting our classical particle at $\vec{x}{ _\perp}=0$ we get:
            \eqn{
            p^i&=\frac{g}{2\pi}T^a \int d^2x_{ _T}\frac{x^i}{x^2_T} \rho^a(\vec{x}_{ _T})\,\,.
            }
            Putting this into the path integral of the MV model, we can extract a probability distribution over~$\vec{p}_{ _T}$: 
            \eqn{&\frac{d P(\vec{p}_{ _T})}{d^2 p_{ _T}} = \frac{1}{\mathscr{N}} \int \mathscr{D}[\rho]\exp\left(-\frac{1}{2\mu^2}\int d^2x\rho^a\rho^a\right)\nonumber\\
            	&\times\delta\!^{(2)}\!\left(\vec{p}_{ _T}-\frac{g}{2\pi}T^a \int d^2x\frac{\vec{x}}{x^2} \rho^a\right)\nonumber\\
            	&=\frac{1}{\mathscr{N}}\int\frac{d^2\lambda}{(2\pi)^2}e^{i\vec{\lambda}\cdot \vec{p}_{ _T}} \int \mathscr{D}[\rho]\exp\left(-\frac{1}{2\mu^2}\int d^2x\rho^a\rho^a\right.\nonumber\\
            	&\left.-i\frac{g}{2\pi}T^a \int d^2x\frac{\vec{\lambda}\cdot\vec{x}}{x^2} \rho^a\right)\label{Pt-probability}
            }
            where \eq{\mathscr{N}=\int \mathscr{D}[\rho]\,e^{-\frac{1}{2\mu^2}\int d^2x\rho^a\rho^a}}
            and the subscript $T$ (which stands for ``transverse") has been omitted for brevity. We complete the square in the argument of the right-most exponential in \eqref{Pt-probability} and then shift the integration variable $\rho$, giving
            
            \eqn{&\frac{d P(\vec{p}_{ _T})}{d^2 p_{ _T}} = \frac{1}{\mathscr{N}}\int\frac{d^2\lambda}{(2\pi)^2} \exp\left(-\frac{mp_0}{2}\vec{\lambda}^2+i\vec{\lambda}\cdot \vec{p}_{ _T} \right)\mathscr{N}\nonumber\\
            	&=\frac{1}{(2\pi)^2} \int d^2\lambda e^{-\frac{1}{2}mp_0\vec{\lambda}^2+i\vec{\lambda}\cdot \vec{p}_{ _T}}\nonumber\\
            	&=\frac{1}{2\pi} \frac{1}{mp_0}\exp\left(-\frac{p_{ _T}^2}{2mp_0}\right) \label{Ptsq-prob}
            }
            with $mp_0=\frac{(gT^a)(gT^a)\mu^2\ln\frac{Q_s}{\Lambda}}{4\pi}$. We have made use of the following fact:
            \eq{\int d^2x\frac{x^ix^j}{x^4}&=\frac{1}{2}\delta^{ij}\int d^2x \frac{1}{x_T^2} \\
            	&=\delta^{ij}\pi\ln{\frac{x_{\text{\tiny{IR}}}}{x_{\text{\tiny{UV}}}}} \,\,
            }
            where $x_{\text{\tiny{UV}}}=1/Q_s$ and infrared cut-off $x_{\text{\tiny{IR}}}$ is the QCD scale $\Lambda_{\text{\tiny{QCD}}}$. 
            
            Doing the angular integration over \eqref{Ptsq-prob}, we get the folowing probability distribution over $p_{\text{T}}^2$
            \eqn{&\frac{d P( p_{ _T}^2)}{d p_{ _T}^2}= \frac{1}{2mp_0}\exp\left(-\frac{p_{ _T}^2}{2mp_0}\right)\nonumber\\
            	&= \frac{2\pi}{(gT^a)(gT^a)\mu^2\ln\frac{Q_s}{\Lambda} }\exp\left(-\frac{2\pi p_T^2}{(gT^a)(gT^a)\mu^2\ln\frac{Q_s}{\Lambda} }\right)\,\,.\label{pt-dist}
            }
             This is an exponential distribution in the variable $p_{ _T}^2$ with the following expectation value 
             \eqn{\langle p_{ _T}^2\rangle=2mp_0=\frac{(gT^a)(gT^a)\mu^2\ln\frac{Q_s}{\Lambda}}{2\pi}\,\,.
             }
            This also determines the distribution of $p^z$ because the equations of motion tell us that $p^-$ is unchanging \cite{Kajantie:2019hft,Kajantie:2019nse}, so 
            
            so we get the following probability distribution over $p^z$
            \eqn{\frac{dP(p^z)}{dp^z}&=\frac{dp_{ _T}^2}{dp^z} \frac{d P( p_{ _T}^2)}{d p_{ _T}^2}\nonumber\\
            	&=2m\frac{1}{2mp_0}\exp\left(-\frac{2mp^z}{2mp_0}\right)\nonumber\\
            	&=\frac{1}{p_0}e^{-\frac{p^z}{p_0}}\,\,.
            }

\section{Derivation of the matter distribution\label{app:Deriv_matter}}

			 The goal is to derive the one-particle distribution for the system of particles described in Section \ref{Matt-dist}. That is, an initially uniform one-dimensional array of particles that is struck by a sheet of colored glass. Let us consider first, the case of a single particle at rest at the origin for all time. The corresponding distribution is obviously given by 
			\eqn{f(x,v)=\frac{dN}{dvdz}&=\delta(z)\delta(v)\,\,,
			}	
		where $v=v^z$ and $x=(t,z)$. We have chosen to work with velocity instead of momentum for the moment, since it makes equations less cumbersome. We will revert to $f(x,p)$ at the end of the derivation.
		
			Next we consider particle sittting at the origin at rest for $t<0$ that is struck and picks up velocity $v'$ at $t=0$ and moves freely from then on. In this case we get
			\eqn{ f(x,v)=\theta(-t)\delta(z)\delta(v)+\theta(t)\delta\left(z-vt\right)\delta(v-v')\, .
		}
				
			We now introduce a velocity distribution by imagining $N_0$ particles sittting at the origin at rest for $t<0$. They then are struck at $t=0$, picking up a range of velocities $v_1',v_2'\dots$, with a  fraction $\rho(v_i')$ of them picking up velocity $v_i'$. They move freely from then on, in which case we get the following distribution
				\eqn{f(x,v)&=N_0\theta(t)\sum\limits_{i}\delta\left(z-vt\right)\delta(v-v_i')\rho(v_i')\nonumber\\
					&+N_0\theta(-t)\delta(z)\delta(v)\,\,,
			}
		with $\,\sum\limits_{i}\rho(v_i')=1$. Passing to the continuum, ${v'_i\rightarrow v'}$, we get
		\eqn{\therefore f(x,v)&=N_0\theta(t)\int_0^1\! dv'\delta\left(z-vt\right)\delta(v-v')\rho(v')\nonumber\\
			&+N_0\theta(-t)\delta(z)\delta(v)\nonumber\\
			&=N_0\theta(t)\delta\left(z-vt\right)\rho(v)+N_0\theta(-t)\delta(z)\delta(v)\,.
		}
		Now we step away from the origin and consider $N_j$ particles with position $z=z_j$ sitting at rest for $t<t_j$ . They are struck at $t=t_j$ by a light-like sheet of coloured glass, picking up a range of velocities ${0<v'<1}$ weighted by a distribution $\rho(v')$ moving freely thereafter. We now get the following distribution
		\eqn{&f(x,v)= N_j\theta(t-t_j)\delta\left(z-z_j-v(t-t_j)\right)\rho(v)\nonumber\\
			&+N_j\theta(t_j-t)\delta(z-z_j)\delta(v) \nonumber\\
			&=N_j\theta(t-z_j)\delta\left(z-z_j-v(t-z_j)\right)\rho(v) \nonumber\\
			&+ N_j\theta(z_j-t)\delta(z-z_j)\delta(v)\,\,.\label{dist-j}
		}
		with 
		\eq{\int_0^1 dv\rho(v)=1\,\,.
		}
		Finally, we consider a chain of these particles, and the corresponding distribution entails summing Eq.~\eqref{dist-j} over the positions of the particles $z_j$, giving 
		\eqn{f(x,v)&=\sum\limits_{j}N_j\theta(t-z_j)\delta\left(z-z_j-v(t-z_j)\right)\rho(v) \nonumber 
			\\&+\sum\limits_{j}N_j\theta(z_j-t)\delta(z-z_j)\delta(v)\,.
		}
		Henceforth, the total number of particles will be denoted ${N=\sum_{j}N_j}$. We can write $N_j$ terms of a line density $N_j=\lambda(z_j)\Delta \,z$, and passing to the continuum for a chain of length $R$ we get:
		\eqn{&f(x,v)=\int_0^R dz'\lambda(z')\theta(z'-t)\delta(z-z')\delta(v\nonumber)\\
			      &+\int_0^R dz' \lambda(z')\theta(t-z')\delta\left(z-z'-v(t-z')\right)\rho(v)\nonumber\\
			      &=\lambda(z)\theta(z-t)\theta(z)\theta(R-z)\delta(v)\nonumber\\
			      &+ \theta(z-vt)\theta((1-v)R-(z-vt))\lambda\left(\frac{z-vt}{1-v}\right)\nonumber\\
			      &\times\theta\left(\frac{t(1-v)-z+vt)}{1-v}\right)\frac{\rho(v)}{1-v}\,\,.
		}
		 We assume a uniform initial distribution, ${\lambda=N/R}$, therefore we get
		 \eqn{&f(x,v)=\frac{N}{R}\theta(z-t)\theta(z)\theta(R-z)\delta(v)\nonumber\\
		 	&+ \frac{N}{R}\theta(t-z)\theta(z-vt)\theta(v(t-R)+R-z)\frac{\rho(v)}{1-v}\,\,.\label{f-x-v}
		 }
		Now we have built up a single-particle distribution  $f(x,v)$ in 1+1 dimensions for arbitrary velocity distribution $\rho(v)$. Let's take a look at the specific $\rho(v)$ resulting from the MV model. To derive this, we use the probability distribution in \eqref{pt-dist} even though it applies to full 3+1 D motion; we will just focus on the $z$-component. Using Eq.\eqref{p-after}, we have
		  	\eq{&p^0=\frac{p^++p^-}{\sqrt{2}}=\frac{1}{\sqrt{2}}\left(\frac{m^2+p_{\text{\tiny{T}}}^2}{\sqrt{2}m}+\frac{m}{\sqrt{2}} \right)\\
		  	&=\frac{p_{\text{\tiny{T}}}^2}{2m}+m=p^z+m=p+m\\
		  	&\therefore v=v^z=\frac{p^z}{p^0}=\frac{p}{p+m}
		  }
		and
		\eqn{&p=\frac{v}{1-v} m\,\,,\\
			&\therefore\frac{dp}{dv}=\frac{m}{(1-v)^2}=\frac{(p+m)^2}{m}
		}
		This allows us to change from the velocity distribution, $\rho(v)$, to momentum a distribution
		\eqn{&\rho(v)=\frac{d\text{P}(v)}{dv}=\frac{dp}{dv}\frac{d\text{P}(p)}{dp}\nonumber\\
			&=  \frac{(p+m)^2}{m}\frac{1}{p_0}e^{-\frac{p}{p_0}}\label{rho-v-to-dProb}
		} 
		where
		\eqn{p_0=\frac{ g^2T^2\mu^2\ln\frac{Q_s}{\Lambda}}{4\pi m}\,\,.
		}
		We have dropped the superscript in $p^z$, and we'll do so henceforth unless it causes confusion. Finally we write the one-particle distribution in terms of $p$ with the help of Eqs.\eqref{f-x-v} and \eqref{rho-v-to-dProb}
		\eqn{&f(x,p)=\frac{dN_{\text{\tiny{matter}}}}{dzdp}=\frac{dv}{dp}\frac{dN}{dz dv}\nonumber\\
			&=\frac{N}{R}\left\lbrace\theta(z-t)\theta(z)\theta(R-z)\delta\left(\frac{p}{p+m}\right)\frac{m}{(p+m)^2}\right.\nonumber\\
			&+\theta(t-z)\theta\left(z-\frac{p\,t}{p+m}\right)\theta\left(\frac{p(t-R)}{p+m}+R-z\right)\nonumber\\
			&\left.\times \frac{p+m}{mp_{_0}}e^{-\frac{p}{p_{_0}}}\right\rbrace\,\,.	
		}

	\section{Derivation of radiation distribution\label{App:RadDeriv}}

\subsection{Number density with formation time and beam rapidity\label{app:RadDerivztTauform}}

	Following the same processes as in the text, we may add a formation time of the form	
		\begin{equation}
			N_0\left(1-\e^{-\tau/\tau_0}\right)\overset{\tau\gg\tau_0}{\longrightarrow}N_0,
		\end{equation}
		
	as well as a beam rapidity $ y_b $.  We will take the characteristic formation time to go like $ \sfrac{1}{Q_s} $, and note that the above expression holds for the proper time, so that $ \tau=t \,\frac{1}{\gamma} =t\sqrt{1-v^2}$.  The phase-space density of the produced particles radiated from a single matter particle at the origin is then be given by
		\begin{align}\label{eq:rad1part2}
			f^{0}(z,t,y)=F\,\T{t}\delta(z-t\tanh y)\left(1-\exp\{-Q_s \sqrt{t^2-z^2}\}\right)
		\end{align}
	
	We now consider a single nucleon at rest, but with an arbitrary position given by $ z=z_i $ and $ t=t_i $.  We may then write
		\begin{align}
			f^{z_i}(z,t,y)=F\,\T{t-t_i}\delta(z-z_i-(t-t_i)\tanh y)\left(1-\exp\left\{-Q_s \sqrt{(t-t_i)^2-(z-z_i)^2}\right\}\right).
		\end{align}
	
	Since we would like to impose a beam rapidity $ y_b $ such that $ v_b =\tanh y_b $, we also have that $ z_i=t_i v_b $, so that
		\begin{align}
			f^{z_i}(z,t,y)=F\,\T{t-\frac{z_i}{v_b}}\delta\left(z-z_i-(t-\frac{z_i}{v_b})\tanh y\right)\left(1-\exp\left\{-\frac{Q_s}{v_b} \sqrt{(tv_b-z_i)^2}-v_b^2(z-z_i)^2\right\}\right).
		\end{align}

	We suppose now that there are $ N_i $ nucleons at each $ z_i $ s.t. $ \sum_i\Delta z = R $ and $ \sum_i N_i=N $.  In the limit $ \Delta z \rightarrow 0 $, equivalent to the limit of a uniform line density of matter particles of $ \sfrac{N}{R} $, we have that
		\begin{align}
			f(z,t,y)	
				=&\int_{0}^{R}\,dz'\frac{F\,N}{R}\,\T{t-\frac{z'}{v_b}}\delta\left(z'\left(\frac{\tanh y}{vb}-1\right)+z-t \tanh y\right)\times\nonumber\\
				&\times\left(1-\exp\left\{-\frac{Q_s}{v_b} \sqrt{(tv_b-z')^2-v_b^2(z-z')^2}\right\}\right).
		\end{align}
	
	As in the case of the matter distribution, we realise here that the integral over $ z' $ has two possibilities: One in which $ z<R $, where the upper limit on the $ z' $-integral should be $ z $, and one in which $ R<z $ where the upper limit on the $ z' $-integral should be $ R $:
		\begin{align}
			f(z,t,y)	
				=\frac{F\,N}{R}\T{tv_b-z}\T{z-t\tanh y}\Big[
					\T{R-z}I^{\rad}_{\text{f},\tau_f}(z)
					+\T{z-R}I^{\rad}_{\text{f},\tau_f}(R)\Big],
		\end{align}
		
	where 
		\begin{align}\label{eq:Iradf}
			I^{\rad}_{\text{f},\tau_f}	(a)
				=&\int_{0}^{a}\,dz'\T{R-z}\T{t-\frac{z'}{v_b}}\delta\left(z'\left(\frac{\tanh y}{vb}-1\right)+z-t \tanh y\right)\nonumber\\
				&\times\left(1-\exp\left\{-\frac{Q_s}{v_b} \sqrt{(tv_b-z')^2-v_b^2(z-z')^2}\right\}\right)\nonumber\\
				=&\frac{v_b\T{tv_b-z}}{v_b-\tanh y}\T{z-t\tanh y}\T{\tanh y \left(t-\frac{a}{v_b}\right)+a-z}
					\left(1-\exp\left\{-\frac{Q_s(z-tv_b)}{\tanh y-v_b}\sech y\right\}\right).
		\end{align}
		
	The theta functions in \cref{eq:Iradf} will determine the integration limits in the integrals that follow.  The number density is then given by
		\begin{align}
			\frac{dN^{\rad}}{dz}
				=&\int_{0}^{\infty}dy \,f(z,t,y)\nonumber\\
				=&\frac{F\,N}{R}\T{tv_b-z}\int_{0}^{\infty}dy\,\T{z-t\tanh y}\Big[
					\T{R-z}I^{\rad}_{\text{f},\tau_f}(z)+\T{z-R}I^{\rad}_{\text{f},\tau_f}(R)\Big]\nonumber\\
				=&\frac{F\,N}{R}\T{tv_b-z}\T{z-t\tanh y}\left[
					\T{R-z}I^{\rad}_{\text{N},\tau_f}(0)
					+\T{z-R}I^{\rad}_{\text{N},\tau_f}\left(\frac{z-R}{t-R}\right)\right],
		\end{align}
	
	where
		\begin{align}
			I^{\rad}_{\text{N},\tau_f}(a)
				&=\int_{\tanh^{-1}(a)}^{\tanh^{-1}(\sfrac{z}{t})}dy\,
					\frac{v_b}{v_b-\tanh y}\left(1-\exp\left\{-\frac{Q_s(z-tv_b)}{\tanh y-v_b}\sech y\right\}\right).
		\end{align}
		
	Lastly, the total energy density may be computed, once again making use of the argument presented in \cref{sec:dEdEta} that $ E(y)=Q_s\cosh y $:
		\begin{align}
			\frac{dE^{\rad}}{dz}
				&=\frac{Q_s}{R}\,F\,N\T{tv_b-z}\T{z-t\tanh y}\left[
					\T{R-z}I^{\rad}_{\text{E},\tau_f}(0)
					+\T{z-R}I^{\rad}_{\text{E},\tau_f}\left(\frac{z-R}{t-R}\right)\right],
		\end{align}
	
	where
		\begin{align}
			I^{\rad}_{\text{E},\tau_f}(a)
				&=\int_{\tanh^{-1}(a)}^{\tanh^{-1}(\sfrac{z}{t})}dy\,
					\frac{v_b \cosh y}{v_b-\tanh y}\left(1-\exp\left\{-\frac{Q_s(z-tv_b)}{\tanh y-v_b}\sech y\right\}\right).
		\end{align}

\subsection{Radiation average velocity}\label{app:radAvgV}
	
	Will present here the derivation of the radiation without using a phase-space density.  
	
	\paragraph{Single nucleon at origin.}
		We start by thinking about a single particle of matter, say a  quark, at the origin, radiating a single particle of radiation, perhaps a gluon with velocity $ v $ at time $ t=0 $.  That radiated gluon then has a position at a later time $ t $ given classically by $ z=tv $.  Therefore, at a given time $ (z,t) $, there is exactly one particle, and it has position $ z=tv $, and velocity $ v $.
		
\begin{figure}
			\centering
			\begin{subfigure}{0.3\linewidth}
				\centering
				\includegraphics[scale=1.]{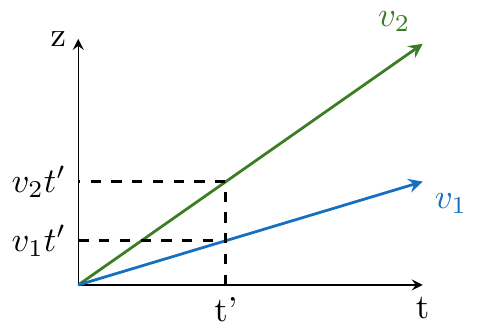}
				\caption{\label{fig:twoFromOrigin}}
			\end{subfigure}
			\begin{subfigure}{0.3\linewidth}
				\centering
				\includegraphics[scale=1.]{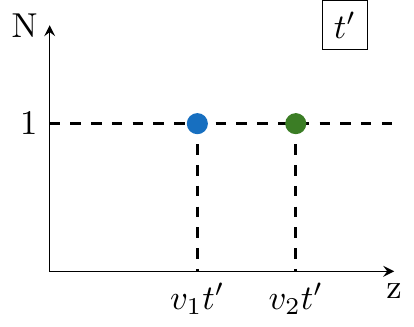}
				\caption{\label{fig:twoFromOriginN}}
			\end{subfigure}
			\begin{subfigure}{0.3\linewidth}
				\centering
				\includegraphics[scale=1.]{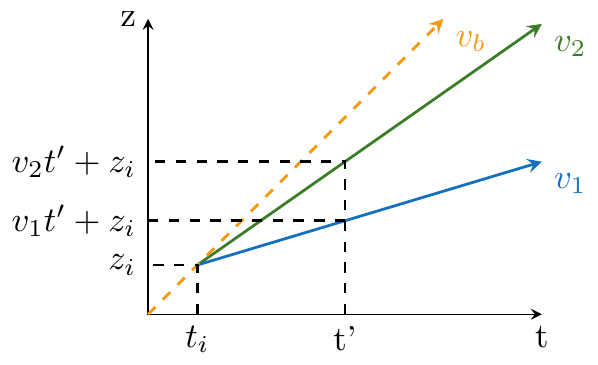}
				\caption{\label{fig:twoFromZi}}
			\end{subfigure}
			\caption{(a)-(b)Two gluons radiated from a quark at the origin, with velocities $ v_1 $ and $ v_2 $. (c) Two gluons radiated from a quark at position $(z_i,\sfrac{z_i}{vb})  $}
		\end{figure}

		If the single quark at the origin radiated two gluons, each with a distinct velocity, say $ v_1 $ and $ v_2 $, then there are two particles in total at a later time $ t $, but they have different locations $ z_1 =v_1 t$ and $ z_2 =v_2 t$.  This situation is illustrated in \cref{fig:twoFromOrigin,fig:twoFromOriginN}.  As an exercise, we may ask what the average velocity is of a particle at a given $(z,t)  $.  Of course, in the current setup there will only ever by at most one gluon at $ (z,t) $ and the average velocity will therefore be the velocity of that particle.  The only particle that will ever have arrived at $ (z,t) $ is precisely the particle that was radiated from the origin with velocity $ v=\sfrac{z}{t} $.  We therefore have that the average velocity (or simply the velocity in this case) is given by		
		\begin{equation}
		\langle v \rangle = 
		\begin{cases}
		v_1		&\mbox{if\,\,} \frac{z}{t}=v_1\\
		v_2		&\mbox{if\,\,} \frac{z}{t}=v_2\\
		0		&\mbox{else.} 
		\end{cases}
		\end{equation}
		
		In fact, even if the quark at the origin radiates $ N $ gluons with a uniformly distributed velocity, the average velocity at $ (z,t) $ will always be $ \sfrac{z}{t} $.  This must of course also be true even if multiple gluons are emitted at a given velocity.
		
		\paragraph{Single nucleon at $ z_i $.} 
		Suppose now that the two gluons are radiated from a quark situated at an arbitrary position $ z_i $.  At this point it is also convenient to introduce a beam rapidity so that the beam velocity will be called $ v_b $.  This will mean that a quark sitting at position $ z_i $ will radiate gluons at time $ t_i = \sfrac{z_i}{v_b} $.  The only particles that will arrive at a given point $ (z,t) $ are those radiated with precisely the correct velocity to allow them to travel to this point.  That is, any particle emitted at point $ z_i $ at a time $ t_i = \sfrac{z_i}{v_b} $ must be emitted with a velocity.
		\begin{equation}
		v=\frac{z-z_i}{t-\frac{z_i}{v_b}}.
		\end{equation}
		
		This situation is illustrated in \cref{fig:twoFromZi}.

		The average velocity of particles at $ (z,t) $ that were emitted from $ (z_i,t_i) $ is then
		\begin{equation}
		\langle v \rangle
		=\Theta(z-z_i)\Theta(t-t_i)\Theta\left(v_b-\frac{z-z_i}{t-\sfrac{z_i}{}v_b}\right)\frac{z-z_i}{t-\frac{z_i}{v_b}},
		\end{equation}
		and the average rapidity is simply, by definition of the momentum-space rapidity
		\begin{align}
		\langle y \rangle	
		=&\frac{1}{2}\log\left\{\frac{t+z-z_i(1+\sfrac{1}{v_b})}{t-z-z_i(1-\sfrac{1}{v_b})}\right\}\times\nonumber\\
		&\times	\Theta(z-z_i)\Theta\left(t-\frac{z_i}{v_b}\right)\Theta\left(t \,v_b-z\right).
		\end{align}
	
\begin{figure}
			\centering
			\includegraphics[scale=1.5]{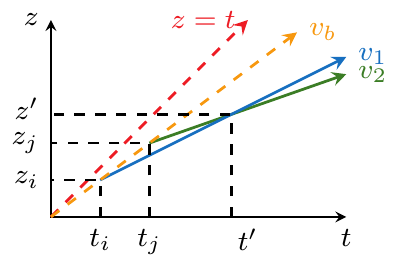}
			\caption{Gluons radiated from different nuclei with different velocities arrive at the same point in $ (z,t) $ and their velocities are to be averaged.\label{fig:zt3}}
		\end{figure}

	\paragraph{Multiple nucleons}
		The difficulty arises from multiple quarks at different positions radiating multiple gluons at different velocities.  In this situation there are multiple gluons at position $ (z,t) $ with various different velocities that one would like to average.  This situation is illustrated in \cref{fig:zt3}.
		
		If multiple gluons are emitted by each nucleon, we require some kind of distribution of their velocities or rapidities, $ F(y) $.  We will assume that this distribution is flat and is appropriately normalized so that
		\begin{equation}
		F(y)=\frac{F}{y_b}.
		\end{equation}
		
		We will also assume that the nucleons are distributed uniformly between $ z=0 $ and $ z=R $, so that, if the nucleons are placed at distances $ \Delta z_i $ apart for $ \Delta zi = \sfrac{R}{n} $, there are $ N_i=\sfrac{N}{n} $ nucleons in each interval.  The average velocity is then
		\begin{align}
		\langle v\rangle
		&=\sum_{i=1}^{n}\frac{y_b}{N_i F n}\times\frac{F N_i}{y_b}\frac{z-z_i}{t-\sfrac{z_i}{v_b}}\\
		&=\frac{1}{n}\sum_{i=1}^{n}\frac{z-z_i}{t-\sfrac{z_i}{v_b}}\\
		&\overset{n\rightarrow\infty}{\rightarrow}\frac{1}{h}\int_{0}^{h}dz'\frac{z-z'}{t-\sfrac{z'}{v_b}},
		\end{align}
		
		where we made the substitution $ \sfrac{1}{n}=\sfrac{\Delta z_i}{R} $, and $ h =\min\{z,R\}$ so that only the correct nucleons contribute to the average.	This integral is straightforward to compute.  Defining		
		\begin{align}
		I^{v}(a)
		&=\int_{0}^{a}dz'\Theta(z-z')\Theta(t v_b-z')\frac{z-z'}{t-\sfrac{z'}{v_b}},
		\end{align}  
		
		we can write
		\begin{align}
		\langle v \rangle
		=&\Theta(t v_b-z)\Bigg[\frac{1}{z}\Theta(R-z)I^{v}(z)+\frac{1}{R}\Theta(z-R)I^{v}(R)\Bigg].
		\end{align}
		
		To get the average rapidity we need only do a slight redefintion:
		\begin{align}
		I^{y}(a)
		&=\int_{0}^{a}dz'\Theta(z-z')\Theta(t v_b-z')\half\ln\left\{\frac{1+\frac{z-z'}{t-\sfrac{z'}{v_b}}}{1-\frac{z-z'}{t-\sfrac{z'}{v_b}}}\right\},
		\end{align} 
		
		So that the average rapidity is
		\begin{align}
		\langle y \rangle
		=&\Theta(t v_b-z)\Bigg[\frac{1}{z}\Theta(R-z)I^{y}(z)+\frac{1}{R}\Theta(z-R)I^{y}(R)\Bigg].
		\end{align}

		It is also possible to introduce a formation time by introducing a factor of $ (1-\exp\{\sfrac{\tau}{\tau_f}\}) $, which becomes
		\begin{equation}
		\left(1-\exp\left\{-Q_s(t-t_i)\sqrt{1-\left(\frac{z-z_i}{t-t_i}\right)^2}\right\}\right).
		\end{equation}
		
		The process described above may then be used with only minor modifications to $ I^{v}(a) $ and $ I^{y}(a) $:
		\begin{align}
		I^{v}_{\tau_f}(a)
		=	&\int_{0}^{a}dz'\Theta(z-z')\Theta(t v_b-z')\frac{z-z'}{t-\sfrac{z'}{v_b}} \times\nonumber\\
		&\times	\left(1-\exp\left\{-Q_s(t-\sfrac{z'}{v_b})\sqrt{1-\left(\frac{z-z'}{t-\sfrac{z'}{v_b}}\right)^2}\right\}\right),\\
		I^{y}_{\tau_f}(a)
		&=\int_{0}^{a}dz'\Theta(z-z')\Theta(t v_b-z')
		\half\ln\left\{\frac{1+\frac{z-z'}{t-\sfrac{z'}{v_b}}}{1-\frac{z-z'}{t-\sfrac{z'}{v_b}}}\right\}\times\nonumber\\
		&\times	\left(1-\exp\left\{-Q_s(t-\sfrac{z'}{v_b})\sqrt{1-\left(\frac{z-z'}{t-\sfrac{z'}{v_b}}\right)^2}\right\}\right),		
		\end{align}

\section{Relativistic coordinates and boosts}{}\label{app:DeriveEtaTau}	

    \subsection{Transformation to ($\eta,\tau$)-space}
    	We would like to transform from $ (z,t) $-space to $ (\eta,\tau) $-space, where $ \eta=\half\ln\frac{t+z}{t-z} $ and $ \tau=\sqrt{t^2-z^2} $,  with inverse transformations $ z=\half\tau(\e^{\eta}-\e^{-\eta}) $ and $ t=\half\tau(\e^{\eta}+\e^{-\eta}) $. We consider the distributions in this paper to be to be a probability density function in $ z $ but not in $ t $, which is to say that they are the distributions in $ z $ at a given, or constant, $ t $.  Note therefore that
    		\begin{align}
    		dN	&=\frac{\partial N}{\partial z}dz+\frac{\partial N}{\partial t}dt\nonumber\\
    		\Rightarrow	\frac{dN}{d\eta}
    		&=\frac{\partial N}{\partial z}\frac{dz}{d\eta}
    		\end{align}
    	
    	The relevant Jacobian for the coordinate transformation we would like to do is therefore
    		\begin{align}
    		\frac{dz}{d\eta}=\frac{t^2-z^2}{t}= \frac{\tau}{\cosh\eta}.
    		\end{align}
    	
    	A consequence of performing the calculation at $ t = c $, for some constant $ c $, is that one must be careful in $ (\eta,\tau) $-space since $ \tau $ is no longer independent of $ \eta $.  To see this, note that
    		\begin{align}
    		\tau&=\sqrt{c^2-t^2} =\sqrt{c^2-\tau^2\sinh^2\eta}\nonumber\\
    		\Rightarrow
    		\tau&=c\sech\eta.
    		\end{align}
		
	\subsection{Boosting to the center-of-mass frame}
	
	    Since the audience for for this paper primarily does research related to heavy-ion collisions in collider experiments, we present in this appendix a pedagogical exploration of comparisons between the fragmentation region that the present work is performed in, and the rest-frame of the center of mass (CoM) in which many standard heavy-ion collision calculations are performed.
	    
	    The first important point to make is that, generally, those working in the CoM frame work in the limit that the colliding nuclei are infinitely thin.  This limit implies that the nuclei are moving asymptotically close to the speed of light and that the collision of the two nuclei happens on asymptotically small time-scales.  In order to be able to make a comparison to the present work, one must relax this assumption some-what and consider perhaps nuclei with a thickness of $\epsilon$, so that the collision occurs from $t=-\epsilon$ to $t=\epsilon$.
	        \begin{figure}
	            \centering
	            \includegraphics{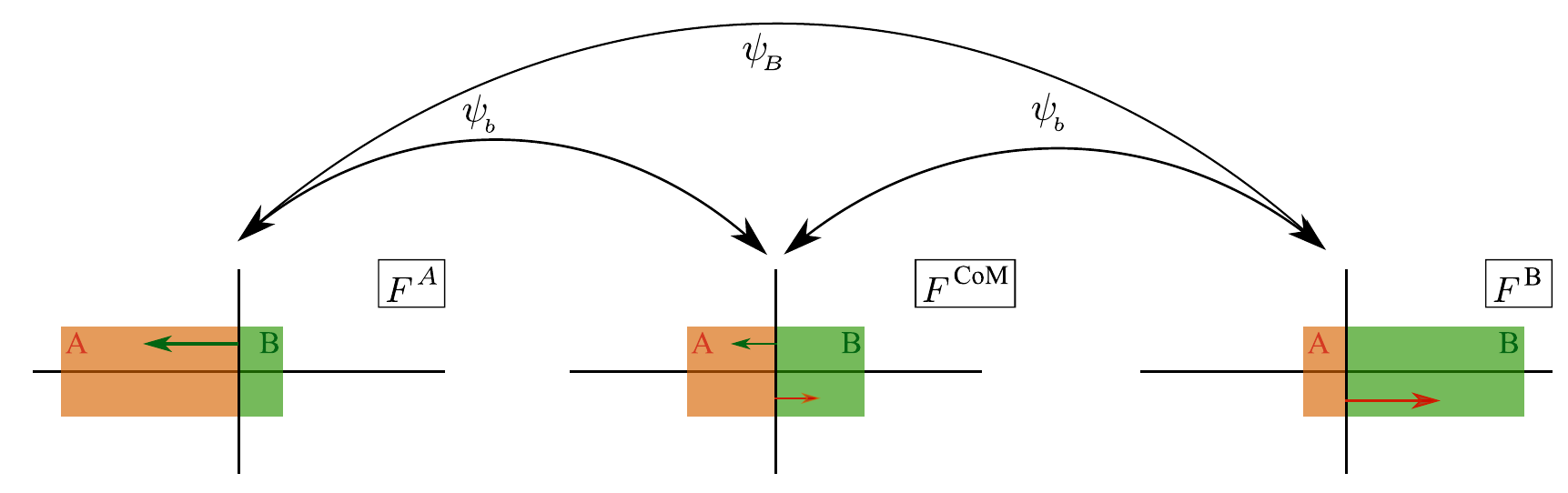}
	            \caption{The start of the collision in three different frames. Boosting from the rest-frame of one nucleus, to the center of mass frame, and on to the rest frame of the other nucleus.}
	            \label{fig:frames}
	        \end{figure}
	    
	    More rigorously, we consider a nucleus $A$ with a length $R_A$ in the rest-frame of $A$ (hereinafter $F^A$). In $F^A$, the left-moving nucleus, nucleus $B$, is boosted from its rest-frame by a velocity which is the relativistic addition of two beam velocities.  That is, the velocity $v_B$ of $B$ in $F^A$, and rapidity $\psi_B$, are given by
	        \begin{align}
	            v_B     &=\frac{2v_b}{1+v_b^2},\\
	            \psi_B  &=\half\ln\left\vert\frac{1+v_B}{1-v_B}\right\vert,
	        \end{align}
	   where $v_b$ is the beam velocity in the CoM frame (hereinafter $F^{CoM}$). We choose to take $\psi_B$ to be a positive number so that we may boost to a forward-moving frame by $[z^+,z^-]\rightarrow[z^+e^{\psi_B},z^-e^{-\psi_B}]$, and to a backward moving frame by  $[z^+,z^-]\rightarrow[z^+e^{-\psi_B},z^-e^{+\psi_B}]$, but of course the same is achieved without the absolute value by using a negative velocity.  In the present work, we consider the problem from the rest-frame of $B$, (hereinafter $F^B$), in which nucleus $A$ is the right-moving projectile.  See \cref{fig:frames} for a cartoon describing the relative frames.
	   
	        \begin{figure}
    	        \begin{subfigure}{0.5\textwidth}
    	            \centering
    	            \includegraphics[scale=0.5]{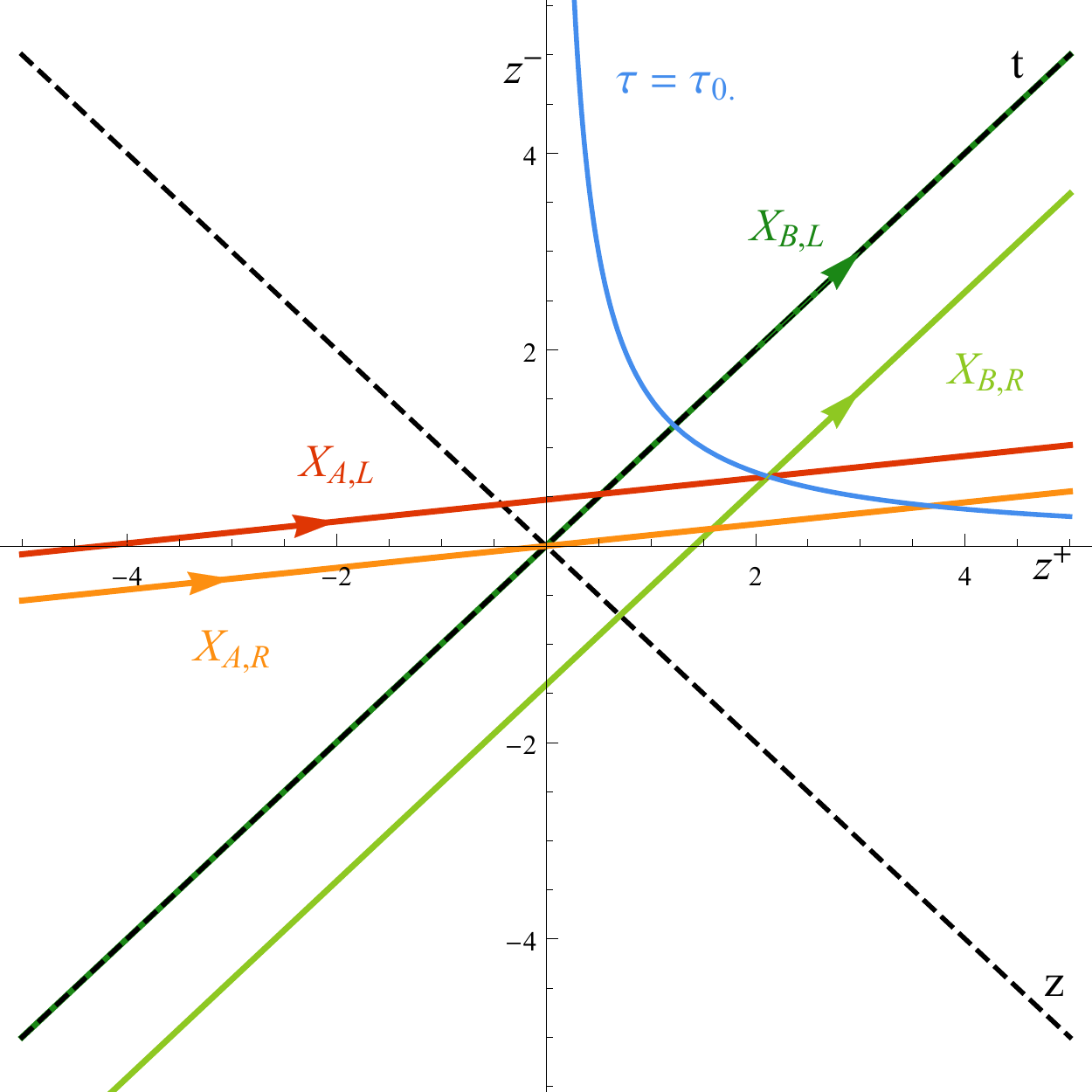}
    	            \caption{In $F^B$}
    	            \label{fig:FA}
    	        \end{subfigure}\begin{subfigure}{0.5\textwidth}
    	            \centering
    	            \includegraphics[scale=0.5]{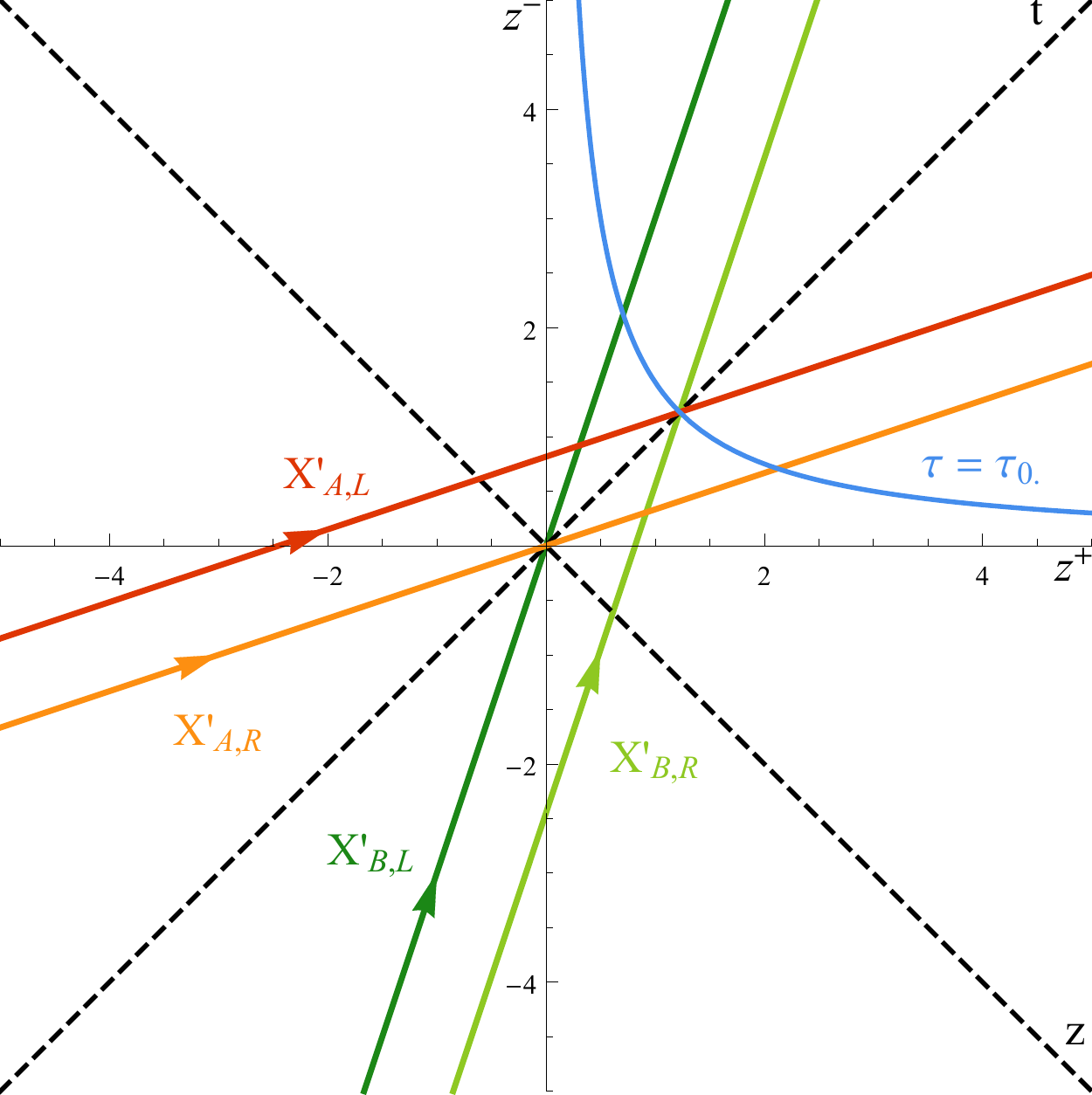}
    	            \caption{In $F^{CoM}$}
    	            \label{fig:FCoM}
    	        \end{subfigure}
    	        \caption{The space-time trajectories of the left (dark) and right (light) edges of nuclei $A$ (green) and $B$ (red) in the rest-frame of (a) nucleus $B$, and (b) the center of mass.  Superimposed (blue) is the line of constant proper time corresponding to the end of the collision.}
	        \end{figure}

	   With these conventions in place we may draw the space-time diagram in \cref{fig:FA}  containing the trajectories of the left- and right-hand edges of nucleus $A$, labeled $X_{A,L}$ and $X_{A,R}$ respectively, as well as the left- and right-hand edges of nucleus $B$, labeled $X_{B,L}$ and $X_{B,R}$ respectively.  For the figure we have chosen an unrealistically low value for $v_b$, at LHC or RHIC energies the boosted lines would be indistinguishable from the $z^+$ and $z^-$ coordinates. We may parametrize the straight lines in \cref{fig:FA} as follows in light-cone coordinates\footnote{The second factor of $e^{\psi_A}$ results from reparameterizing the boosted light-cone vector by shifting the +-component.}:
	   
	        \begin{align}\label{eq:FBparam}
	            X_{A,L} &= [z^+,e^{-2\psi_A}z^++\sqrt{2}\,R_A e^{-\psi_A}]  \nonumber\\
	            X_{A,R} &= [z^+,z^+e^{-2\psi_A}] \nonumber\\
	            X_{B,L} &= [z^+,z^+]  \nonumber\\
	            X_{B,R} &= [z^+,z^+-\sqrt{2} \,R_B]. 
	        \end{align}
	   
	   We define the origin in frames $F^{CoM}$ and $F^A$ such that the collision ``starts'' (the front edges of the two nuclei coincide) at $(t=0,z=0)$ in both frames. The nuclei \emph{separate} when the left-hand edge of nucleus $A$ coincides with the right-hand edge of nucleus $B$, and in $F^{CoM}$ this separation happens at $(t=t_{0,CoM},z=0)$. This is the time at which calculations of the evolution of the matter produced in heavy-ion collisions are generally started, and the separation event lies on the hyperbola defined by proper time $\tau_0=t_{0,CoM}$.  Some manipulation of the parameterization in \cref{eq:FBparam} will reveal that
	        \begin{equation}
	            \tau_0=\frac{R}{\sinh{\psi_b}}=\frac{R}{v_b}\sqrt{1-v_b^2}
	        \end{equation}
	   
	   It follows that in the limit of infinite beam rapidity, we have $\tau_0\rightarrow0$. One may perform the algebraic exercise of boosting the parameterization in \cref{eq:FBparam} to $F^{CoM}$ and again finding the intersection of $X'_{B,R}$ and $X'_{A,L}$, which must correspond to the same value of $\tau_0$ as it did in $F^B$. Note that in $F^B$ the separation happens at a different point, $z=R$ instead of $z=0$.  An illuminating exercise is to compute the separation time in $F^{B}$ and compare it to $t_{0,CoM}$.  As must be the case, one finds that the separation time in $F^{B}$ (denoted $t_{0,B}$) is the time-dilated equivalent of $t_{0,CoM}$:
	   
	        \begin{align}
	            t_{0,B}     &=\frac{R}{\tanh\psi_b}\label{eq:t0B}\\
	            t_{0,CoM}   &=\frac{R}{\sinh\psi_b}=\frac{R}{\tanh\psi_b}\frac{1}{\cosh\psi_b}
	                            =\frac{t_{0,B}}{\cosh\psi_b}.\label{eq:t0CoM}
	        \end{align}
	        
	    We may now consider the limit in which $v_b\rightarrow\infty$.  We see that the time that it takes for the collision to occur in $F^B$, given by \cref{eq:t0B}, goes to $R$, while the time for the collision to occur in $F^{CoM}$ goes to $0$.  One of the main goals of the present work is precisely to develop an intuition for the physics of a heavy-ion collision that is obscured in the center of mass frame in the limit of very high beam velocities.

\end{document}